\documentclass[aps,prd,twocolumn,superscriptaddress,floatfix,amsmath,amssymb,notitlepage,showpacs,10pt]{revtex4-1}

\usepackage{graphicx}
\usepackage{enumitem}
\usepackage{bbm}
\usepackage{relsize}
\usepackage{xcolor}
\usepackage[caption=false]{subfig}
\usepackage{amsmath}	
\usepackage{amssymb}    
\usepackage[pdftex]{hyperref}
\usepackage[english]{babel}
\allowdisplaybreaks[2]

\begin{document}


\title{Pulse shape optimization for electron-positron production in rotating fields} 

\author{Fran\c{c}ois Fillion-Gourdeau}
\email{francois.fillion@emt.inrs.ca}
\affiliation{Universit\'{e} du Qu\'{e}bec, INRS-\'{E}nergie, Mat\'{e}riaux et T\'{e}l\'{e}communications, Varennes, Qu\'{e}bec, Canada J3X 1S2}
\affiliation{Institute for Quantum Computing, University of Waterloo, Waterloo, Ontario, Canada, N2L 3G1}

\author{Florian Hebenstreit}
\email{hebenstreit@itp.unibe.ch}
\affiliation{Albert Einstein Center, Institut f\"{u}r Theoretische Physik, Universit\"{a}t Bern, 3012 Bern, Switzerland}

\author{Denis Gagnon}
\email{denis.gagnon@emt.inrs.ca}
\affiliation{Universit\'{e} du Qu\'{e}bec, INRS-\'{E}nergie, Mat\'{e}riaux et T\'{e}l\'{e}communications, Varennes, Qu\'{e}bec, Canada J3X 1S2}
\affiliation{Institute for Quantum Computing, University of Waterloo, Waterloo, Ontario, Canada, N2L 3G1}

\author{Steve MacLean}
\email{steve.maclean@emt.inrs.ca}
\affiliation{Universit\'{e} du Qu\'{e}bec, INRS-\'{E}nergie, Mat\'{e}riaux et T\'{e}l\'{e}communications, Varennes, Qu\'{e}bec, Canada J3X 1S2}
\affiliation{Institute for Quantum Computing, University of Waterloo, Waterloo, Ontario, Canada, N2L 3G1}


\begin{abstract}
We optimize the pulse shape and polarization of time-dependent electric fields to maximize the production of electron-positron pairs via strong field quantum electrodynamics processes. The pulse is parametrized in Fourier space by a B-spline polynomial basis, which results in a relatively low-dimensional parameter space while still allowing for a large number of electric field modes. The optimization is performed by using a parallel implementation of the differential evolution, one of the most efficient metaheuristic algorithms. The computational performance of the numerical method and the results on pair production are compared with a local multistart optimization algorithm. These techniques allow us to determine the pulse shape and field polarization that maximize the number of produced pairs in  computationally accessible regimes. 
\end{abstract}

\pacs{12.20.Ds, 11.15.Tk, 02.60.Pn}

\maketitle


\section{Introduction}

The generation of electron-positron pairs from strong classical electric fields has been predicted theoretically decades ago \cite{Sauter:1931zz,Heisenberg:1935qt,PhysRev.82.664} but still eludes a direct experimental verification. Generating electromagnetic radiation with an electric field strength on the order of the Schwinger field $E_{S} := m^{2} c^{3}/ e\hbar \approx 1.3 \times 10^{18}$ V/m ($m$ is the electron mass and $e$ is the electric charge), is the main challenge experimentalists are facing to observe this effect. Since the probability to generate a pair from a constant field $E_{\rm exp}$ is proportional to $P_{S} \sim \exp(-\pi E_{S}/E_{\rm exp})$, the pair density is exponentially suppressed for $E_{\rm exp} < E_{S}$. With current laser technologies, the maximum attainable peak field strength is approximately given by $E_{\rm exp} \sim 10^{13}-10^{14}$ V/m \cite{RevModPhys.78.309}, which is  several
orders of magnitude below the Schwinger field and, hence, results in a minuscule pair production probability ($P_{S} \ll 1$).  

Nevertheless, the latest developments in laser science that aim at increasing the laser intensity, along with new theoretical proposals, have made the experimental observation of the Schwinger mechanism more plausible \cite{PhysRevSTAB.5.031301,PhysRevLett.94.103903,RevModPhys.84.1177}. In this line of research, the development of time-dependent pulses or field configurations that enhance the pair density has been one of the main guiding principles. As a matter of fact, it has been demonstrated numerous times that the pair density depends nonlinearly on the field and is hence very sensitive to its temporal structure \cite{PhysRevLett.102.150404,PhysRevD.82.045007,Abdukerim2013820,PhysRevLett.112.050402,akal2014electron,PhysRevLett.104.250402,PhysRevA.86.032118,PhysRevD.88.045017,PhysRevD.89.085001,PhysRevD.70.053013,PhysRevE.66.016502}. Among the most promising field configurations are those that realize the dynamically assisted Schwinger effect \cite{PhysRevLett.101.130404,PhysRevD.80.111301,PhysRevA.85.033408}, whereby a strong quasi-static electric field is superimposed by weak high-frequency radiation. This increases the pair production rate and reduces the exponential suppression owing to the combination of tunnelling and multiphoton effects. 

Most theoretical calculations that investigate the effect of the laser pulse shape on the pair density have considered homogeneous fields, although some spatial effects have also been studied recently \cite{PhysRevLett.102.080402,PhysRevLett.107.180403,PhysRevLett.110.013002,0953-4075-46-17-175002,PhysRevD.94.065024}. Even for the simpler homogeneous but oscillating electric field with an envelope, it is far from trivial to understand the pair spectrum and to find an optimal configuration due to time-domain quantum interferences. The latter accounts for the intricate ``peak and valley'' structure in the pair spectrum \cite{PhysRevLett.102.150404, Abdukerim2013820,PhysRevD.95.056013}. The phenomenon of quantum interference can be understood as a realization of the Stokes phenomenon \cite{PhysRevLett.104.250402,PhysRevLett.108.030401} or the Landau-Zener-St\"{u}ckelberg interferometry (LZSI) \cite{PhysRevA.86.032118} and renders the pair spectrum extremely sensitive to the field profile \cite{Abdukerim2013820}. 

Recently, it has been proposed to optimize the time profile parameters with optimal control theory to maximize the pair density \cite{PhysRevD.88.045028}. Later, a similar technique was utilized to optimize the rate in some momentum region \cite{Hebenstreit2014189}. The main impetus of this study was to find pulse shapes that facilitate the detection of Schwinger's effect in an experimental investigation. Even with recent advances in laser technologies, the laser pulses will need to be tightly focused to reach the required field strength. This can only be performed with large focusing optics that cover a substantial part of the whole solid angle. As a consequence, it may be challenging to design a particle detector with a large acceptance. In addition, these particle detectors have some given momentum resolution and range. It is then mandatory that pairs are emitted in the direction of the particle detector with the proper energy. These experimental constraints could be theoretically controlled by using optimization techniques.

Following a similar approach based on optimal control, it is also possible to formulate the inverse problem for Schwinger pair production to determine electric field configurations that approximately reproduce a given particle spectrum \cite{Hebenstreit2016336}. The success of this procedure depends on the number of modes that characterize the laser pulse. Generally speaking, the accuracy of the solution improves exponentially as the number of modes is increased. At the same time, the computational cost to find optimal solutions grows polynomially with the size of the search space if the mode amplitudes are chosen as the optimization parameters as in Ref. \cite{Hebenstreit2016336}. This limits the number of modes to $\mathcal{O}(10)$, which is unrealistically small for short laser pulses that exhibit a broadband spectrum. 
Another limitation of the optimization studies previously performed in Refs.~\cite{PhysRevD.88.045028,Hebenstreit2014189,Hebenstreit2016336} is the assumption that the electric field is spatially homogeneous and linearly polarized.
Although this assumption is well-justified for an e-dipole field in the vicinity of the focal spot or at the antinodes of two counter-propagating standing waves, recent investigations have demonstrated that an elliptic or circular polarization of the electric field can have substantial influence on the properties of produced particles \cite{PhysRevD.89.085001, PhysRevD.91.045016,PhysRevD.92.085001}. For instance, it has been shown that a circular polarized field results in a ring structure in the spectrum of created particles and leads to a non-trivial spin polarization \cite{PhysRevD.91.125026,PhysRevD.92.085001,PhysRevD.93.025014,2016arXiv160102468K}. 

The main goal of this article is to go beyond the aforementioned limitations by improving the method outlined in Refs. \cite{PhysRevD.88.045028, Hebenstreit2014189,Hebenstreit2016336} in various respects: 
\begin{itemize}[leftmargin=.13in]
\item \textit{The pulse parametrization}:
We parametrize the pulse in Fourier space by using a polynomial basis expansion. Once the spectrum is parametrized in this polynomial basis, a larger number of modes can be used. If the pulse spectrum is smooth enough, the number of necessary parameters to completely characterize the pulse can be reduced significantly compared to a direct optimization of the spectral weights and phases. 

\item \textit{The optimization technique:}
We perform a comparison of two optimization strategies, namely the commonly used multistart local search, and a more general approach based on metaheuristics. Metaheuristics are well suited for large scale optimization problems as they can usually find good solutions with less computational resources than other methods, especially when the parameter space has many local minima \cite{blum2003metaheuristics}. On the other hand, local search algorithms usually have a faster rate of convergence if the parameter space is convex.

\item \textit{The possible field configurations:} 
We allow for arbitrary ellipticities in the field configurations instead of restricting calculations to the linearly polarized case.
\end{itemize}
We note that pulse shape optimization problems are also encountered in the control of wave-packets in molecular systems \cite{PhysRevA.37.4950}, in harmonic generation in atomic physics \cite{PhysRevA.64.021403,PhysRevA.69.041802,PhysRevA.90.023855, PhysRevLett.86.5458}, in ablation problems \cite{Turbis01012014}, or in the design of high-fidelity quantum gates \cite{dong2016learning,2017arXiv170304165G}.
Thus, the optimization strategies presented in this article may find application in several physical systems.

This article is organized as follows. In Sec.~\ref{sec:can_quant} we briefly describe two methods to evaluate the spectrum of produced electron-positron pairs from strong electric fields.
In Sec.~\ref{sec:model} we introduce the piecewise polynomial basis expansion (B-splines) which is used to represent the electric fields under consideration.
In Sec.~\ref{sec:optimization} we outline two optimization methods that are utilized in the current study: a local multistart optimization algorithm as well as a population-based metaheuristics. We discuss our numerical results on pulse optimization in Sec.~\ref{sec:results} and conclude in Sec.~\ref{sec:conclu}.

\section{Pair production in a strong homogeneous field}
\label{sec:can_quant}

In this section we briefly review electron-positron pair production in a time-dependent homogeneous classical electric field. In particular, we present two independent techniques. 
The first one is based on the solution of the Dirac equation in momentum space. It is adapted from the formulation given in Refs. \cite{Gelis20161, PhysRevA.86.032118,PhysRevB.92.035401,fillion2016numerical}. The second one is a generalization of the quantum kinetic equation for linearly polarized fields \cite{Schmidt:1998vi,PhysRevD.60.116011,PhysRevD.82.105026} to two-dimensional rotating electric fields. Both techniques yield equivalent results and will be utilized in subsequent optimization calculations. The performance of these approaches is compared in Sec.~\ref{sec:perf_num}.
Throughout we use QED rationalized units in which $c=\hbar=m=1$ and $e=\sqrt{4\pi \alpha}$, such that $eE_{S}=1$. Moreover, we choose the temporal-axial gauge in which $A_0(t,\mathbf{x})=0$.

\subsection{Pair production from the Dirac equation}
\label{sec:DE}
 
The electron-positron phase space density in a homogeneous external field is computed from the field induced transitions between negative and positive energy states.
Therefore, the leading order contribution to the spin-summed electron phase space density $f(t,\mathbf{p})$ can be written as \cite{PhysRevB.92.035401}
\begin{align}
f(t_{f},\mathbf{p})  = \sum_{s=1,2} 
\frac{1}{2\omega^{\mathrm{out}}(\mathbf{p}) 2\omega^{\mathrm{in}}(\mathbf{p}) }
 \left|  u^{\mathrm{out} \dagger}_{s}(\mathbf{p}) \psi_{s}(t_{f},\mathbf{p}) \right|^{2},
\label{eq:pair_prod_homo}
\end{align}
where $s$ is the spin index, $\omega^{\mathrm{in,out}}(\mathbf{p})$ are the asymptotic energies, $u^{\mathrm{out} }_{s}(\mathbf{p})$ is a positive energy spinor and $\psi_{s}(t_{f},\mathbf{p})$ is a relativistic wave function. The superscripts ``$\mathrm{in},\mathrm{out}$'' implies that the variable is evaluated in regions where the electric field vanishes, before and after the field is applied.  Eq.~\eqref{eq:pair_prod_homo} is based on the assumption that the electric field vanishes at asymptotic times, i.e. $\left. \mathbf{E}(t)\right|_{t \in [-\infty,t_{i}] \cup [t_{f},\infty]} = 0$. In turn, the vector potential, related to the electric field as usual by $\mathbf{E}(t) = - \partial_{t}\mathbf{A}(t)$, may be non-vanishing in these temporal regions depending on the chosen gauge. Here, we choose gauges where the vector potential is spatially constant but time-dependent and given by $\left. \mathbf{A}(t)\right|_{t \in [-\infty,t_{i}]} = \mathbf{A}^{\mathrm{in}}$ and $\left. \mathbf{A}(t)\right|_{t \in [t_{f},\infty]} = \mathbf{A}^{\mathrm{out}}$. 

At this point, it is convenient to introduce the kinematic momentum and the single particle energy according to
\begin{subequations}
\begin{align}
\label{eq:kin_mom}
\mathbf{P}(t) &:=  \mathbf{p} - e \mathbf{A}(t) \ , \\
\omega(\mathbf{p},t) &:= \sqrt{ \mathbf{P}^{2}(t) + m^{2} } \ .
\end{align}
\end{subequations}
The asymptotic energies are simply given by $\omega^{\mathrm{in,out}}(\mathbf{p}) := \omega(\mathbf{p},t_{i,f})$. 
Moreover, the adiabatic free spinors can be written as
\begin{subequations}
\begin{align}
\label{eq:free_spin1}
u_{s}(t,\mathbf{p}) &= \frac{1}{\sqrt{\omega(\mathbf{p},t) + m }}
\begin{bmatrix}
[\omega(\mathbf{p},t) + m ] \xi_{s} \\
[\boldsymbol{\sigma} \cdot \mathbf{P}(t)] \xi_{s}  
\end{bmatrix} ,\\  
v_{s}(t,-\mathbf{p})& = \frac{1}{\sqrt{\omega(\mathbf{p},t) + m }}
\begin{bmatrix}
-[\boldsymbol{\sigma} \cdot \mathbf{P}(t)] \eta_{s}  \\
[\omega(\mathbf{p},t) + m ] \eta_{s} 
\end{bmatrix},
\end{align}
\end{subequations}
where the bi-spinors are defined as $\xi_{1} = \eta_{2} = [1,0]^{\mathsmaller T}$ and $\xi_{2} = \eta_{1} = [0,1]^{\mathsmaller T}$. It can be verified that the spinors obey the usual orthogonality relations $u^{\dagger}_{s}(t,\mathbf{p})v_{s}(t,-\mathbf{p}) = 0$ which ensures that the pair density vanishes for free propagation without electric field. 

In Eq.~\eqref{eq:pair_prod_homo}, the superscript ``$\mathrm{in},\mathrm{out}$'' indicate that the spinors are evaluated at times $t_{i}$ and $t_{f}$, respectively ($u_{s}^{\mathrm{out}}(\mathbf{p}) := u_{s}(t_{f},\mathbf{p})$ and $v_{s}^{\mathrm{in}}(\mathbf{p}): = v_{s}(t_{i},\mathbf{p})$).
Accordingly, the momentum space wave function is given by 
\begin{align}
\label{eq:wf_init}
\psi_{s}(t_{f},\mathbf{p}) = U_{\mathbf{p}}(t_{f},t_{i}) v_{s}^{\mathrm{in}}(-\mathbf{p}) \ ,
\end{align}
where $U_{\mathbf{p}}$ is the evolution operator in momentum space. It evolves an initial negative energy free spinor $v_{s}^{\mathrm{in}}$ from the initial time $t_{i}$ to the final time $t_{f}$ according to the momentum space Dirac equation
\begin{align}\label{eq:dirac_eq_mom}
i\partial_{t}\psi(t,\mathbf{p}) =  \left[ \boldsymbol{\alpha} \cdot   \left[ \mathbf{p}  - e\mathbf{A}(t) \right] + \beta m \right] \psi(t,\mathbf{p}) \ .
\end{align} 
The Dirac matrices are chosen in the usual Dirac representation and thus, are given by $\boldsymbol{\alpha} = \sigma_{x} \otimes \boldsymbol{\sigma}$ and $\beta = \sigma_{z} \otimes \mathbb{I}_{2}$, with Pauli matrices $\boldsymbol{\sigma}$. The time evolution of the wave function is performed by solving the Dirac equation \eqref{eq:dirac_eq_mom} for initial negative energy states of momentum $\mathbf{p}$, selected from a given range. To this end, we use a simple split-operator method with a second order convergence which was developed in previous studies \cite{PhysRevA.86.032118,PhysRevB.92.035401,fillion2016numerical}. The time step is adjusted to reach convergence of the solution.

We  conclude
this section by noting that the conservation of charge and momentum allows for a relation between the electron and positron phase space density. The latter is obtained by the substitution $\mathbf{p} \rightarrow -\mathbf{p}$ in the electronic $f(t,\mathbf{p})$. Finally, the total pair density generated by the electric field is obtained by integrating $f(t,\mathbf{p})$ over all momenta.

\subsection{Pair production from quantum kinetic theory}
\label{sec:QKE}

Equivalently, the evolution of a Dirac field under the influence of an external vector potential $\mathbf{A}(t,\mathbf{x})$ can be suitably described using the Dirac-Heisenberg-Wigner (DHW) phase space approach \cite{PhysRevD.44.1825}. For a spatially homogeneous vector potential $\mathbf{A}(t)$ with $\mathbf{E}(t)=-\partial_t \mathbf{A}(t)$, this formalism appears as a linear system of PDEs for 10 nontrivial Wigner components $\mathbbm{w}=[\,\mathbbm{s},\vec{\mathbbm{v}},\vec{\mathbbm{a}},\vec{\mathbbm{t}}_1]^{\mathsmaller T}$,
\begin{equation}
 \label{eq:DHW}
 \left[\partial_t+e\,\mathbf{E}(t)\cdot\nabla_\mathbf{p}\right]\mathbbm{w}(t,\mathbf{p}) = \mathcal{M}(\mathbf{p})\,\mathbbm{w}(t,\mathbf{p}) \ 
\end{equation}
with
\begin{equation}
 \mathcal{M}(\mathbf{p})=\begin{pmatrix}0&0&0&2\mathbf{p}^{\mathsmaller T}\\0&0&-2\mathbf{p}\times&-2m\,\mathbbm{1}\\
 0&-2\mathbf{p}\times&0&0 \\ -2\mathbf{p}&2m\,\mathbbm{1}&0&0\end{pmatrix} \ ,
\end{equation}
and nontrivial initial conditions $\mathbbm{s}(t_i,\mathbf{p})=-2m/\omega(\mathbf{p})$ and $\vec{\mathbbm{v}}(t_i,\mathbf{p})=-2\mathbf{p}/\omega(\mathbf{p})$ with $\omega(\mathbf{p})=\sqrt{\mathbf{p}^2+m^2}$. We note that for linearly polarized fields $\mathbf{E}(t)=E_3(t)\mathbf{e}_3$, the PDE system can be reduced to a 3-dimensional ODE system via the method of characteristics  \cite{PhysRevD.44.1825,PhysRevD.82.105026}. 

In the following, we consider a two-dimensional electric field and parameterize it as $\mathbf{E}(t)=E_2(t)\mathbf{e}_2+E_3(t)\mathbf{e}_3$. It has been noted in Ref.~\cite{PhysRevD.89.085001} that there exists a possible redundancy in the 10-dimensional PDE systems in similarity to linearly polarized fields. However, this redundancy was not lifted and the full system was solved. As we will now show, it is in fact possible to reduce the system to a subset of only six equations by a suitable choice of basis. To this end, we proceed as in \cite{PhysRevD.82.105026} and apply the method of characteristics to transform Eq.~\eqref{eq:DHW} in the form
\begin{equation}
 \partial_t\mathbbm{w}(t,\mathbf{P})=\mathcal{M}(\mathbf{P})\,\mathbbm{w}(t,\mathbf{P}) \ ,
\end{equation}
with the kinematic momentum $\mathbf{P}$ as defined in Eq.~\eqref{eq:kin_mom}. We then seek an appropriate basis to span the Wigner components
\begin{equation}
 \mathbbm{w}(t,\mathbf{P}):=-2\sum_{i=1}^{10}\chi_i(t,\mathbf{P})\,\mathbbm{e}_i(t,\mathbf{P}) \ ,
\end{equation}
with normalized basis vectors $\mathbbm{e}_i(t,\mathbf{P})$ and expansion coefficients $\chi_i(t,\mathbf{P})$. For convenience, the first basis vector is chosen such that it encodes the nontrivial initial conditions $\chi_1(t_i,\mathbf{P})=1$ whereas $\chi_{i=2\ldots 10}(t_i,\mathbf{P})=0$. Moreover, we introduce the following quantities
\begin{subequations}
\begin{align}
 \mathcal{X}(t,\mathbf{P})&:=\omega^2(\mathbf{P})(\mathbf{E}\cdot\mathbf{E})-(\mathbf{E}\cdot\mathbf{P})^2 \ , \\
 \mathcal{Y}(t,\mathbf{P})&:=\omega^2(\mathbf{P})(\mathbf{E}\cdot\dot{\mathbf{E}})-(\mathbf{E}\cdot\mathbf{P})(\dot{\mathbf{E}}\cdot\mathbf{P}) \ ,
\end{align}
\end{subequations}
with $\dot{\mathbf{E}}:=\partial_t\mathbf{E}$. In fact, the choice of normalized basis vectors
\begin{subequations}
\begin{align}
 \mathbbm{e}_1&:=\frac{1}{\omega}[m,\mathbf{P},0,0\,]^{\mathsmaller T} \\
 \mathbbm{e}_2&:=\frac{1}{\omega\sqrt{\mathcal{X}}}[m(\mathbf{E}\cdot\mathbf{P}),(\mathbf{E}\cdot\mathbf{P})\mathbf{P}-\omega^2(\mathbf{P})\mathbf{E},0,0\,]^{\mathsmaller T} \\
 \mathbbm{e}_3&:=-\frac{1}{\sqrt{\mathcal{X}}}[0,0,\mathbf{E}\times\mathbf{P},m\,\mathbf{E}\,]^{\mathsmaller T} \ , \\
 \mathbbm{e}_4&:=\frac{1}{\epsilon_\perp\sqrt{\mathcal{X}}}\notag \\
 &\mkern+30mu[-m(\mathbf{e}_1\times\mathbf{P})\cdot\mathbf{E},\mathbf{E}\times(m^2\mathbf{e}_1+p_1\mathbf{P}),0,0\,]^{\mathsmaller T} \\
 \mathbbm{e}_5&:=\frac{1}{\epsilon_\perp\omega\sqrt{\mathcal{X}}(\mathbf{E}\times\dot{\mathbf{E}})\cdot\mathbf{e}_1} \notag \\
 &\mkern+30mu[0,0,(\mathcal{X}\dot{\mathbf{E}}-\mathcal{Y}\mathbf{E})\times\mathbf{P},m(\mathcal{X}\dot{\mathbf{E}}-\mathcal{Y}\mathbf{E})^{\mathsmaller T}  \\
 \mathbbm{e}_6&:=-\frac{1}{\epsilon_\perp\omega}[0,0,m^2\mathbf{e}_1+p_1\mathbf{P},m\mathbf{e}_1\times\mathbf{P}\,]^{\mathsmaller T}  
\end{align}
\end{subequations}
with $\epsilon_\perp=\sqrt{m^2+p_1^2}$ results in a closed subsystem of equations. By defining the two auxiliary quantities
\begin{subequations}
\begin{align}
 Q(t,\mathbf{P})&:=\frac{e\sqrt{\mathcal{X}(t,\mathbf{P})}}{\omega^2(\mathbf{P})} \ , \\
 R(t,\mathbf{P})&:=\frac{\epsilon_\perp\omega(\mathbf{P})(\mathbf{E}\times\dot{\mathbf{E}})\cdot\mathbf{e}_1}{\mathcal{X} (t,\mathbf{P})} \ , 
\end{align}
\end{subequations}
we obtain
\begin{subequations}
\begin{align}
 \mathcal{M}\,\mathbbm{e}_1&=0 \mkern+70mu&\partial_t\mathbbm{e}_1&=-Q\,\mathbbm{e}_2 \ , \\
 \mathcal{M}\,\mathbbm{e}_2&=2\omega\,\mathbbm{e}_3 &\partial_t\mathbbm{e}_2&=Q\,\mathbbm{e}_1+R\,\mathbbm{e}_4 \ , \\
 \mathcal{M}\,\mathbbm{e}_3&=-2\omega\,\mathbbm{e}_2 &\partial_t\mathbbm{e}_3&=-R\,\mathbbm{e}_5 \ , \\
 \mathcal{M}\,\mathbbm{e}_4&=-2\omega\,\mathbbm{e}_5 &\partial_t\mathbbm{e}_4&=-R\,\mathbbm{e}_2 \ , \\
 \mathcal{M}\,\mathbbm{e}_5&=2\omega\,\mathbbm{e}_4 &\partial_t\mathbbm{e}_5&=Q\,\mathbbm{e}_6+R\, \mathbbm{e}_3 \ , \\
 \mathcal{M}\,\mathbbm{e}_6&=0 &\partial_t\mathbbm{e}_6&=-Q\,\mathbbm{e}_5 \ .
\end{align}
\end{subequations}
This means that the subset $\chi_{i=1\ldots 6}(t,\mathbf{P})$ fully characterizes the fermion dynamics in the presence of a two-dimensional electric field. Accordingly, we obtain the following ODE system by equating the coefficients:
\begin{equation}
 \label{eq:qke}
 \partial_t \vec{\chi}(t,\mathbf{P})=\mathcal{N}(t,\mathbf{P})\vec{\chi}(t,\mathbf{P}) \ ,
\end{equation}
which is governed by the skew-symmetric matrix
\begin{equation}
 \label{eq:matrix}
 \mathcal{N}(t,\mathbf{P}):=\begin{pmatrix} 0 & -Q & 0 & 0 & 0 & 0 \\ 
                              Q & 0 & -2\omega & R & 0 & 0 \\ 
                              0 & 2\omega & 0 & 0 & -R & 0 \\
                              0 & -R & 0 & 0 & 2\omega & 0 \\
                              0 & 0 & R & -2\omega & 0 & Q \\ 
                              0 & 0 & 0 & 0 & -Q & 0 \end{pmatrix} \ ,
\end{equation}
and initial conditions $\chi_1(t_i,\mathbf{P})=1$ and $\chi_{i=2\ldots 6}(t_i,\mathbf{P})=0$.
The single-particle distribution function, which corresponds to the spin-summed pair density, is identified with \cite{PhysRevD.82.105026}
\begin{equation}
 f(t,\mathbf{P})=1-\chi_1(t,\mathbf{P}) \ .
\end{equation}
For linearly polarized fields we note that $R(t,\mathbf{P})=0$. In this special case, Eq.~\eqref{eq:matrix} takes a block-diagonal form, which implies that it suffices to solve a 3-dimensional subsystem, corresponding to the well-known quantum kinetic equation in differential form \cite{PhysRevD.60.116011}. The ODE system can be solved efficiently by using a standard fourth-order Runge-Kutta method. Again, the time step is adjusted to reach convergence.

\section{Temporal field profile}
\label{sec:model}

In this work we consider spatially homogeneous, time-dependent electric fields. Specifically, we will look at two-dimensional field configurations of the form
\begin{equation}
 \mathbf{E}(t) = [0,E_2(t),E_3(t)\,]^{\mathsmaller T} \ .
\end{equation}
Such fields can be generated physically at the antinodes of counter-propagating lasers beams or by using more sophisticated configurations such as the combination of e-dipoles \cite{PhysRevA.86.053836}. They are fully characterized by their spectral density $\tilde{E}_l(\omega)$ and spectral phase $\phi_l(\omega)$, for $l=2,3$. Here, $\omega$ is the angular frequency of the electric field (not to be confused with the relativistic energies defined in Sec.~\ref{sec:DE} and \ref{sec:QKE}) .

Using a polynomial basis expansion, these quantities can be written as 
\begin{subequations}
\begin{align}
\label{eq:coeff_basis_E}
 \tilde{E}_l(\omega) &= \sum_{i=1}^{N_{s}} a^{(E)}_{l,i} B_{i}(\omega) \ , \\
 \label{eq:coeff_basis_phi}
 \phi_l(\omega) &= \sum_{i=1}^{N_{s}} a^{(\phi)}_{l,i} B_{i}(\omega) \ ,
\end{align}
\end{subequations}
where $N_{s}$ is the number of basis elements $B_{i}$ and $a^{(E/\phi)}_{l,i}$ are the corresponding expansion coefficients. The latter will be used as fitting parameters over which the optimization is carried out. 

The spectral density and phase are expanded over B-spline basis functions of the polynomial order $k$,
\begin{equation}
 B_{i}(\omega)=b_{i}^{(k)}(\omega) \ .
\end{equation}
A thorough description of these functions can be found in Ref. \cite{0034-4885-64-12-205}. This choice is favored over other polynomial bases as (i) they have compact support, (ii) they are positive definite, and (iii) they are easy to implement. B-splines are fully determined by the polynomial order $k$ and a given knot vector $(\omega_{i})_{i=1,\ldots,N_{s}+k}$ according to the iterative relation \cite{0034-4885-64-12-205,nla.cat-vn991654}
\begin{align}
 b_{i}^{(k)}(\omega)&= \frac{\omega-\omega_{i}}{\omega_{i+k-1} - \omega_{i}}b^{(k-1)}_{i}(\omega) + \frac{\omega_{i+k} -\omega}{\omega_{i+k}-\omega_{i+1}} b^{(k-1)}_{i+1}(\omega) ,
\end{align}
with initial conditions
\begin{equation}
 b_{i}^{(1)}(\omega)=\begin{cases} 1 & \mbox{for} \quad \omega_{i} \leq \omega < \omega_{i+1}\\ 0 & \mbox{otherwise}\end{cases}.
\end{equation}
The number of knots at a given frequency determines the continuity condition at that point. In this work, we use the standard choice with knots of multiplicity $k$ at the endpoints $\omega_{\rm{min}}$ and $\omega_{\rm{max}}$, and simple knots at the interior points \cite{0034-4885-64-12-205}
\begin{align}
\omega_{\rm{min}}  = \omega_{1}  = \ldots & = \omega_{k} < \cdots  \nonumber \\ 
 < \omega_{k+n-1} & = \ldots =\omega_{2k+n-2} = \omega_{\rm{max}} \, ,
\end{align}
where $n$ is the number of breakpoints and $2k+n-2$ is the number of knot points. These two quantities are related to the number of basis function as $N_{s} = n+k-2$. 

The bandwidth of the electric field is fixed by the endpoints $\omega_{\mathrm{min}}$ and $\omega_{\mathrm{max}}$: Outside of this interval, the spectral density and phase are zero. 
Moreover, the standard choice for the knot vector leads to smooth functions on the whole interval except at the endpoints at which discontinuities may occur. As a consequence, the value of the spectral density and phase are not constrained at these points. Finally, in this work we choose equidistant breakpoints on the whole interval, even though this is not mandatory; more points could be added in some frequency ranges if one desired a higher resolution.  

For a given spectral density and phase, the field is determined by sampling the spectrum at equidistant frequencies $\Omega_{j} = (N_{\mathrm{min}} + j)\Delta \omega$, where $\Delta \omega$ is the angular frequency difference between each mode and $j\in[0,j_{\rm{max}}]\subset\mathbb{N}$ results in a an electric field that is periodic with period $T = 2\pi / \Delta \omega $. Using this notation, we have $\omega_{\mathrm{min}} = \Omega_{0}$ and $\omega_{\mathrm{max}} = \Omega_{N}$. Consequently, the electric field components are given by a coherent superposition of $N+1$ modes
\begin{align}
\label{eq:field}
E_{l}(t) &= g(t) \sum_{j=0}^{N} E_{l,j}\cos(\Omega_{j}t - \Phi_{l,j}) \ ,
\end{align}
with the field strength $E_{l,j}:=\tilde{E}_l(\Omega_j)$ and phase $\Phi_{l,j}:=\phi_l(\Omega_j)$ sampled from the B-spline parametrization. Here, we included an envelope function $g(t)$ for two reasons: First, it ensures that the field vanishes smoothly at finite times $t<t_i$ and $t>t_f$.  Without the envelope, the field would not reach zero after one period unless all the phases vanished. Second, $g(t)$ acts as an apodization function which reduces the spectral leakage. The latter is due to the fact that the signal has a finite time extent. By smoothing the field when it turns on and off, the apodization function reduces the spectral components outside of the interval $[\omega_{\mathrm{min}},\omega_{\mathrm{max}}]$, which improves the accuracy of the spectral representation. Many different choices exist for implementing these windowing techniques \cite{heinzel2002spectrum}. Here, we use the simple Hann function which is similar to a laser pulse
\begin{equation}
 \label{eq:hann}
 g(t):=\cos^2(\Omega_T t) \ ,
\end{equation}
with $\Omega_T:= \pi/T$. 

We may define for each mode a dimensionless Keldysh parameter \cite{Keldysh:1964ud,PhysRevD.2.1191} 
\begin{equation}
 \label{eq:keldysh}
 \gamma_{l,i}:=\frac{m \Omega_{i}}{e|E_{l,i}|} \ , 
\end{equation}
which determines whether the electron-positron production mechanism for this mode is dominated by multiphoton absorption $(\gamma_{l,i}\gg 1)$ or nonperturbative tunnelig $(\gamma_{l,i}\ll 1)$. We note, however, that the notion of a Keldysh parameter is only airtight for monochromatic fields whereas its meaning in multiscale problems is less obvious \cite{PhysRevLett.101.130404,Orthaber:2011cm}. In general, an electric field pulse contains low-frequency modes in the tunneling regime which are dynamically assisted by high-frequency modes. In this sense, some of the field configurations considered in this article realize a multimodal generalization of the dynamically assisted pair production mechanism. 

Given the electric field configurations in Eq.~\eqref{eq:field}, we may easily calculate the associated vector potential
\begin{align}
 \label{eq:vector_potential}
A_{l}(t) &=  A_{l,0} - \sum_{j=0}^{N} E_{l,j} F_{l,j}(t) \ ,
\end{align}
with
\begin{align}
F_{l,j}(t)=&  \frac{1}{4} \left( \frac{2\sin (\Omega_{j}t - \Phi_{l,j})}{\Omega_{j}} + \frac{\sin \left[(2\Omega_{T}-\Omega_{j})t + \Phi_{l,j} \right]}{2\Omega_{T}-\Omega_{j}}\right. \nonumber \\
&\left.\mkern+0mu +\frac{\sin \left[(2\Omega_{T}+\Omega_{j})t - \Phi_{l,j} \right]}{2\Omega_{T}+\Omega_{j}}\right) .
\end{align}
The adjustable constant $A_{l,0}$ ensures that the vector potential vanishes at the final time $t_f$.

Finally, we do not allow for arbitrary large field strengths $E_{l,j}$ but rather require that the integrated energy density of the field
\begin{align}
 \label{eq:energy}
 U:=\int_{t_i}^{t_f}{\mkern-10mu dt\,\mathbf{E}^2(t)}
\end{align}
takes on a fixed value. From a physical point of view, this requirement fixes the total flux of energy through the focus while keeping the polarization of the electric field arbitrary. Most notably, this allows different frequency modes to have different polarizations. This constraint also relates to experimental limitations. Typically, the amount of energy in a pulse emitted from a high intensity laser is a fixed parameter, determined by the laser configuration and hardware. 

\section{Optimization techniques}
\label{sec:optimization}

The basic idea behind optimization problems is to tune a set of control parameters to find an optimal value of a certain quantity. In the current study, we are interested in finding an optimal temporal field profile that maximizes the pair production rate. Formally, the optimization problem is defined as
\begin{align}
\tilde{J} = \max_{[\vec{\mathbf{a}}^{(E)},\vec{\mathbf{a}}^{(\phi)}]} J[A] \ ,
\end{align}
where $J[A]: \mathbb{R}^{2N_{s}}\otimes[0,2\pi]^{2N_s} \rightarrow \mathbb{R}$ is the objective functional (observable) whose value implicitly depends on the temporal profile of the vector potential $\mathbf{A}(t)$. In the present study, the latter is parametrized by the set of basis expansion coefficients via Eqs. \eqref{eq:coeff_basis_E} and \eqref{eq:coeff_basis_phi}. This defines our parameter space as
\begin{subequations}
\begin{align}
 \vec{\mathbf{a}}^{(E)}&:=[\vec{a}_2^{(E)},\vec{a}_3^{(E)}]
 \in\mathbb{R}^{2N_s} \ , \\
 \vec{\mathbf{a}}^{(\phi)}&:=[\vec{a}_2^{(\phi)},\vec{a}_3^{(\phi)}]
 \in[0,2\pi]^{2N_s} \ .
\end{align}
\end{subequations}
Here, $\tilde{J}$ is the value of the global maximum of the objective functional in parameter space. The main goal of any optimization method is to find an accurate approximation of $\tilde{J}$ by suitably changing the value of the parameters $\vec{\mathbf{a}}^{(E)}$ and $\vec{\mathbf{a}}^{(\phi)}$. Specifically, we choose as objective functional the pair density integrated over some momentum domain $\mathcal{D}_{\mathbf{p}}$, such that
\begin{equation}
 \label{eq:opt_prob}
 J[A]:=\int_{\mathcal{D}_{\mathbf{p}}}{\mkern-10mu d^3p\,f(t_f,\mathbf{P})} \ ,
\end{equation} 
which is supplemented by the constraint that the integrated energy density of the field as defined in Eq.~\eqref{eq:energy} takes a fixed value $U_{\rm{const}}$.

In this work, we will apply two fundamentally different optimization approaches in order to (i) cross-check and guarantee a proper operation of each of these methods and (ii) compare the efficiency of the different methods for the current problem. On the one hand we use local search algorithms as in Refs.~\cite{PhysRevD.88.045028,Hebenstreit2014189,Hebenstreit2016336}. Since these algorithms are designed to find local extrema of the objective functional, the optimization has to be repeated a certain number of times with random initial conditions in order to increase the probability that the global maximum is among the local maxima that have been found. This approach is especially useful if the objective functional is convex or the number of local maxima is small. On the other hand, we also use a population based metaheuristics, which explores many maxima of the objective function as it does not stick to a given basin of attraction and allows to scan a large part of the parameter space.

\subsection{Local search and multistart method}
\label{sec:opt_local}

To solve the optimization problem as defined in Eq.~\eqref{eq:opt_prob}, we may employ the approach outlined in Refs.~\cite{PhysRevD.88.045028,Hebenstreit2014189}. Accordingly, there are two types of constraints: First, the integrated pair density is defined in terms of the single-particle distribution function $f(t_f,\mathbf{P})=1-\chi_1(t_f,\mathbf{P})$ (as discussed in Sec.~\ref{sec:QKE}), which is obtained as the solution of the ODE system
\begin{equation}
 \label{eq:eom2}
 \partial_t \vec{\chi}(t,\mathbf{P})-\mathcal{N}(t,\mathbf{P})\vec{\chi}(t,\mathbf{P})=:\vec{e}(t,\mathbf{P})=0 \ , 
\end{equation}
with initial conditions $\chi_1(t_i,\mathbf{P})=1$ and $\chi_{i=2,\ldots,6}(t_i,\mathbf{P})=0$. Second,  the integrated energy density takes a fixed value $U_{\mathrm{const}}$, which is dealt with by introducing a penalty method based on the constraint functional
\begin{equation}
 \mathcal{C}[A]:=U_{\mathrm{const}}-U=U_{\mathrm{const}}-\int_{t_i}^{t_f}{\mkern-10mu dt\,\mathbf{E}^2(t)} \ ,
\end{equation}
which measures deviations from $U_{\mathrm{const}}$. The augmented Lagrangian, which is employed to turn the full constrained optimization problem into an unconstrained one, is then defined according to
\begin{align}
 \mathcal{L}:= - J[A] - \lambda\,\mathcal{C}[A] + \frac{1}{2\mu}\,\mathcal{C}[A]^2 + \sum_{i=1}^{6}\langle e_i,\lambda_i \rangle_{\Omega} \ ,
\end{align}
with Lagrange multiplier fields $\lambda_i(t,\mathbf{P})$ and a penalty parameter $\mu>0$. Here, $\langle \cdot , \cdot\rangle_\Omega$ denotes the $L^2$ inner product on $\Omega=\mathbbm{R}^3\times[t_i,t_f]$. The third term quadratic in $\mathcal{C}[A]$ penalizes constraint violations while a Lagrange multiplier $\lambda$ is included to avoid ill-conditioning of the optimization problem \cite{Nocedal06}.  

In this formulation, local maxima of the particle number correspond to local minima of the Lagrangian. In fact, the variation of the augmented Lagrangian with respect to $\vec{\chi}(t,\mathbf{p})$ yields the adjoint equations
\begin{equation}
 \label{eq:adj}
 \partial_t\vec{\lambda}(t,\mathbf{P})-\mathcal{N}(t,\mathbf{P})\vec{\lambda}(t,\mathbf{P})=0 \ , 
\end{equation}
which have to fulfill the final conditions $\lambda_i(t_f,\mathbf{P})=-1$ and $\lambda_{i=2,\ldots,6}(t_f,\mathbf{P})=0$. Accordingly, while solving Eq.~\eqref{eq:eom2} forward in time, the adjoint equations Eq.~\eqref{eq:adj} are solved backwards in time. From these solutions, the gradient of the Lagrangian with respect to the field parameters $\vec{\mathbf{a}}=[\vec{\mathbf{a}}^{(E)},\vec{\mathbf{a}}^{(\phi)}]$ can be calculated as   
\begin{align}
\label{eq:grad}
 \nabla\mathcal{L}=\sum_{i,j=1}^{6}\langle \chi_i,[\nabla\mathcal{N}]_{ij}\lambda_j \rangle_{\Omega} -\left(\lambda-\frac{\mathcal{C}[A]}{\mu}\right)\nabla\mathcal{C}[A] \ ,
\end{align}
where $\nabla\mathcal{N}$ denotes the elementwise gradient of Eq.~\eqref{eq:matrix}. The latter can be evaluated analytically because the vector potential is explicitly known. In Eq. \eqref{eq:grad}, the first term is responsible for maximizing the particle number while the second term tries to minimize energy constraint violations. The stationary of the gradient, $\nabla\mathcal{L}=0$, is a necessary condition for a local extremum of the augmented Lagrangian. 

To take full advantage of the gradient $\nabla\mathcal{L}$, which determines the local descent direction of the Lagrangian in parameter space, we employ a local optimization algorithm along with a multistart method. The general outline of the algorithm is as follows (for further algorithmic details we refer to Ref.~\cite{Nocedal06}):
\begin{enumerate}[leftmargin=.18in]
\item[1.]{(Initialization) Choose a random initial configuration $\vec{\mathbf{a}}_{0}^{(0)}=[\vec{\mathbf{a}}^{(E)}, \vec{\mathbf{a}}^{(\phi)}]$ along with initial values for the Lagrange multiplier $\lambda^{(0)}=0$ and the penalty parameter $\mu^{(0)}>0$. The value of $\mu^{(0)}$ determines how severely constraint violations are penalized in the first iteration $[l=0]$.}
\item[2.]{(Iterative minimization) At the $l$--th iteration with Lagrange multiplier $\lambda^{(l)}$ and penalty parameter $\mu^{(l)}$, search the local minimizer of $\mathcal{L}(\lambda^{(l)},\mu^{(l)})$ iteratively:
\begin{equation}
 \vec{\mathbf{a}}_{k+1}^{(l)}=\vec{\mathbf{a}}_k^{(l)}+\alpha_k\vec{\mathbf{d}}_k \quad , \quad k\in\mathbbm{N}_0 \ .
 \end{equation}
We calculate the local search directions $\vec{\mathbf{d}}_k$ according to the Broyden-Fletcher-Goldfarb-Shanno  (BFGS) algorithm, and viable step sizes $\alpha_k$ are found via an inexact line search that fulfills the strong Wolfe conditions.}
\item[3.]{(Increased penalization) After converging in the $l$--th iteration to a local mi\-nimum of the Lagrangian  $\widetilde{\mathcal{L}}(\lambda^{(l)},\mu^{(l)})$ for field parameters $\widetilde{\mathbf{a}}^{(l)}$, update the Lagrange multiplier and penalty parameter
\begin{subequations}
\begin{align}
 \lambda^{(l+1)}&=\lambda^{(l)}-\mathcal{C}[\widetilde{\mathbf{a}}^{(l)}]/\mu^{(l)} \ , \\
 \mu^{(l+1)}&=\xi\,\mu^{(l)} \ ,
\end{align}
\end{subequations}
with $0<\xi<1$. This choice guarantees that constraint violations are more severly penalized in the subsequent iteration.}
\item[4.]{(Local minimization) Set the starting point for the $(l+1)$--th iteration according to
\begin{equation}
 \vec{\mathbf{a}}_{0}^{(l+1)}=\widetilde{\mathbf{a}}^{(l)} \ ,
\end{equation}
and repeat the steps $2$ -- $3$ until the constraint is exactly fulfilled, $\mathcal{C}[A]=0$. The corresponding minimum of the augmented Lagrangian is a local solution of the constrained optimization problem.}
\item[5.]{(Global minimization) Repeat the steps $1$ -- $4$ with different initial confi\-gurations (multistart approach) in order to find the global minimum of the augmented Lagrangian.}
\end{enumerate}
The functionality of this local search algorithm is depicted schematically in Fig.~\ref{fig:optimization_algorithms}a.

\subsection{Population based metaheuristics}
\label{sec:opt_meta}

We also employ population-based metaheuristics as an alternative approach to maximize the pair production. This optimization scheme is combined with the Dirac technique as described in Sec.~\ref{sec:DE} for the evaluation of the pair density. In particular, the field parameters $\vec{\mathbf{a}}=[\vec{\mathbf{a}}^{(E)},\vec{\mathbf{a}}^{(\phi)}]$ are optimized by using a parallel version of differential evolution (DE) \cite{Storn1997,price2006differential, das2011differential}. DE is very efficient to solve optimization problems on continuous variables and can be faster than other metaheuristic algorithms \cite{vesterstrom2004comparative}. Here, a variant of DE that goes by the name of self-adaptive differential evolution is used  \cite{1688311}.

\begin{figure}[b]
\raggedright
\includegraphics[width=0.7\columnwidth]{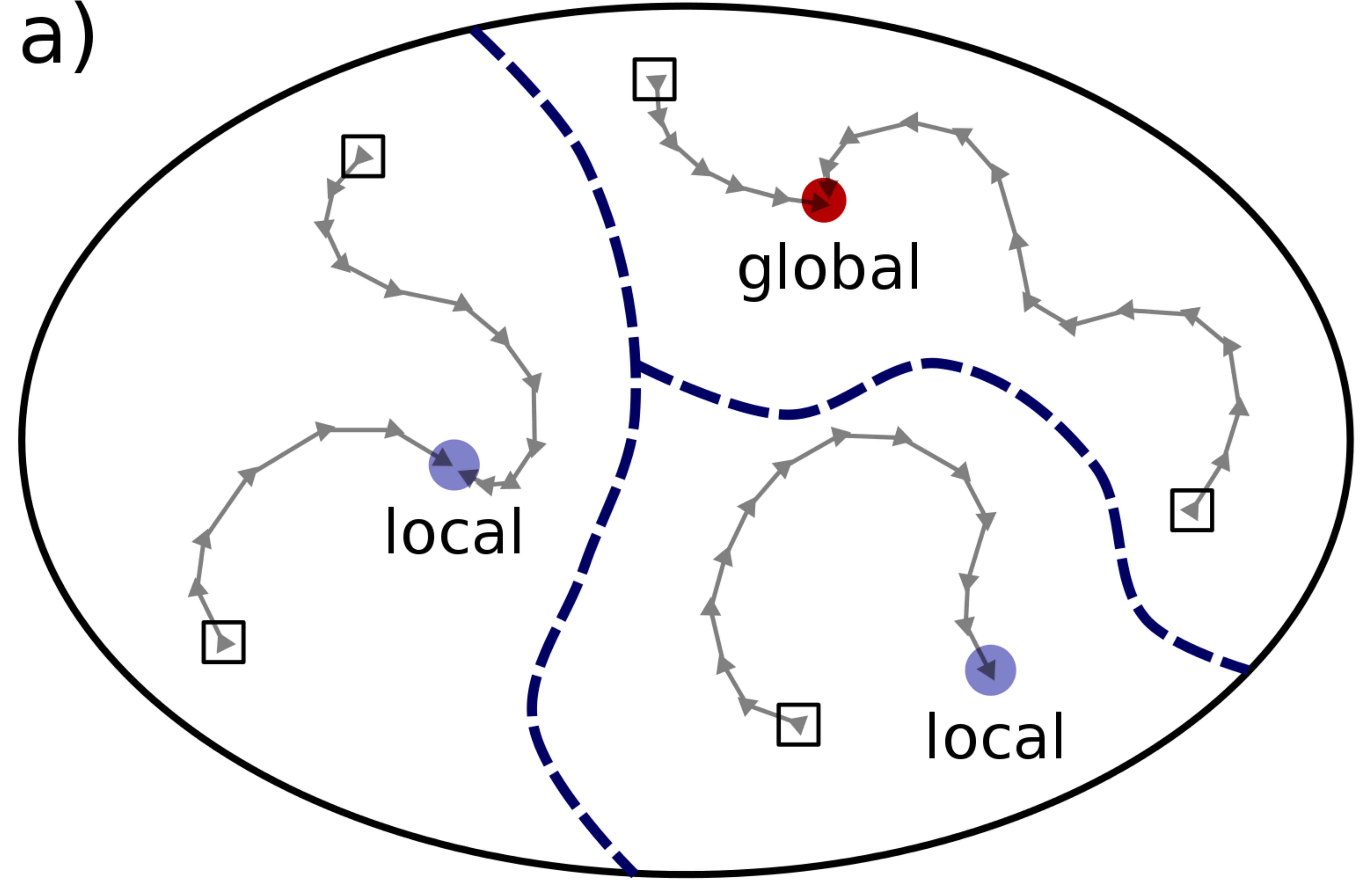} \\
\raggedright
\includegraphics[width=\columnwidth]{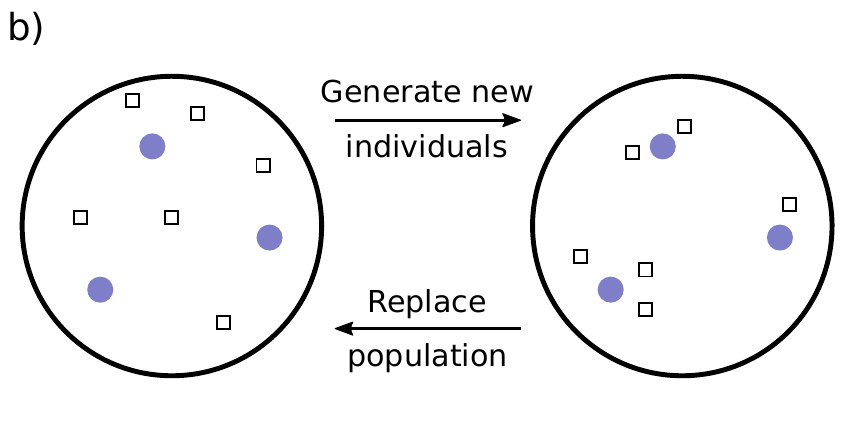}
\caption{Schematic representation of the different optimization schemes. (a) In a local search with multistart we choose different initial conditions in parameter space (black squares). The optimization trajectories then remain in the local basins of attraction (indicated by the dashed lines) and converge towards the corresponding local minima, amongst them also the global minimum. This approach is especially useful if the number of local extrema is small. (b) In population based metaheuristics, different initial conditions or ``individuals'' (black squares) are generated randomly in the parameter space. This population of individuals is iteratively improved using successive applications of mutation, crossover and selection operators. Every generation features individuals successively closer to local minima of the problem. In the specific case of DE, the parameter space is continuous, and the mutation operator is based on vector differences. Inspired by \cite{Talbi2009}.}
\label{fig:optimization_algorithms}
\end{figure}

DE is a population-based metaheuristic where one starts with a number of randomly selected candidate solutions (individuals). Then, individuals are updated using sequential application of mutation, crossover and selection operators. The typical update procedure (DE/rand/1/bin in usual DE notation) can be summarized as follows \cite{1688311, Talbi2009} (see also Fig. \ref{fig:optimization_algorithms}b for a schematic visualization):
\begin{enumerate}[leftmargin=.18in]
 \item[1.]{For each individual $\vec{x}$ in the population, choose three other distinct individuals $\vec{a},\vec{b},\vec{c}$ randomly, also in the population.}
 \item[2.]{(Mutation) Create a vector $\vec{u}$ with components $u_{i} = a_{i} + F(b_{i}-c_{i})$, the so-called mutation operation. }
 \item[3.]{(Crossover) Pick a random integer $R$ from the set $\{1,\cdots,N_{p}\}$, where $N_{p}$ is the size of the parameter space.
		 	For each component $(x_{i})_{i=1,\cdots,N_{p}}$, pick a uniformly distributed random number $r_{i}$ from the interval $[0,1]$. If $r_{i}\leq C_{R}$ or if we look at the $R$'th component ($i=R$), then we set $y_{i} = u_{i}$.
		 	Otherwise, we have $y_{i} = x_{i}$. The condition $i=R$ is included to ensure that at least one of the components of the trial vector $\vec{y}$ is inherited from the mutated vector $\vec{u}$ \cite{Talbi2009}.}
 \item[4.]{(Selection) If $J(\vec{y}) < J(\vec{x})$, replace $\vec{x}$ by $\vec{y}$ in the population.} 
 \item[5.]{Repeat the process from step 1.}
\end{enumerate}
Here, $C_{R} \in [0,1]$ is the crossover probability and $F \in [0,2]$ is the differential weight. They are the only control parameters of the method, along with the number of individuals in the population. In self-adaptive DE, these parameters evolve during the evolution and several mutation schemes are used (besides DE/rand/1/bin). For more details, we refer the reader to Ref. \cite{1688311}.

To ensure that the integrated energy density takes a fixed value $U_{\mathrm{const}}$, we globally renormalize the electric field in every update step. In other words, the field coefficients $\vec{\mathbf{a}}^{(E)}$ represent relative spectral weights. This is different from the local search algorithm in which a penalty method is used.  

In this article, we use the parallel implementation in the Pagmo library \cite{izzo2012pygmo}. The latter implements the Generalized Island model (GIM) parallelization of optimization algorithms \cite{Rucinski2010555,Izzo2012}. This paradigm is useful to execute optimization algorithms on parallel computers with a satisfactory load balancing. According to this principle, the total population is first separated into a number of subpopulations sent to different islands. Then, on each of these islands, the optimization algorithm (DE in our case) is carried out, as usual. In practice, each island is dealt with by a different processor, although this is not mandatory. The main feature of the GIM relates to the migration policy, which allows for the transfer of individuals between different islands, allowing for a mixing of populations. This exchange of information proceeds according to pre-defined migration rules and for a given island topology. The migration rules determines which individuals are migrated and at what frequency, along with the direction of the population transfer. In turn, the topology is defined by a graph type which specifies the connectivity between islands. In this work, a simple ring topology is used as it has demonstrated the best performance when combined with DE \cite{Rucinski2010555}. This topology limits the propagation of the best candidates, which turns out to be beneficial for the DE algorithm. 

\section{Results}
\label{sec:results}

In the following, we study particle production in 1-dimensional (1-D) and 2-dimensional (2-D) electric field configurations. By going beyond the case of linear polarization as investigated in Refs.~\cite{PhysRevD.88.045028, Hebenstreit2014189,Hebenstreit2016336}, we enlarge the parameter space to take into account effects due to the nontrivial polarization.
Specifically, we consider a time-dependent electric pulse as parametrized in Sec.~\ref{sec:model} and we use both the local search algorithm and the population based metaheuristics from Sec.~\ref{sec:optimization} to maximize the pair density. Accordingly, we consider the objective functional as defined in Eq.~\ref{eq:opt_prob}, where the choice of momentum window $\mathcal{D}_{\mathbf{p}}$ is discussed below.

In the 1-D case we optimize the particle number along the field direction, i.e., we disregard the transverse momentum components,
\begin{subequations}
\begin{align}
 \mathbf{E}(t)&=[0,0,E(t)]^{\mathsmaller T} \ , \\
 \mathcal{D}_{\mathbf{p}}&=\{\mathbf{p}\in\mathbb{R}^3|p_1=p_2=0\} \ .
\end{align}
\end{subequations}
Here, the focus lies on the applicability and quality of the polynomial basis expansion using B-splines. We will show that a comparatively small number of basis functions $B_i(\omega)$ suffices to obtain a good approximation of the multimodal electric field. Most important, an increase in the number of basis functions basically does not alter the optimal field configuration and momentum distribution.

In the 2-D case, we consider a momentum sheet in the plane in which the electric field rotates,
\begin{subequations}
\begin{align}
 \mathbf{E}(t)&=[0,E_2(t),E_3(t)]^{\mathsmaller T} \ , \\
 \mathcal{D}_{\mathbf{p}}&=\{\mathbf{p}\in\mathbb{R}^3|p_1=0\} \ .
\end{align}
\end{subequations}
In this case, the emphasis is placed on effects due to the polarization of the electric field, i.e., the phase relation between the field components $E_2(t)$ and $E_3(t)$. Finally, we will use 2-D results to compare the performance of the numerical methods (Dirac equation vs. quantum kinetic theory) and the used optimization algorithms (local search vs. metaheuristics).

\subsection{Optimization for linear polarization in 1-D}
\label{sec:opt_tot_rate_1D}

\begin{table}[t!]
\begin{tabular}{lc}
\hline \hline
Pulse characteristics & Value (QED units) \\
\hline
Minimum frequency ($\nu_{\mathrm{min}} = \omega_{\mathrm{min}}/2\pi$)& 0.001\\
Maximum frequency ($\nu_{\mathrm{max}} = \omega_{\mathrm{max}}/2\pi$)& 0.01\\
Sampling frequency ($\Delta \nu$) & 0.0002 \\
Pulse length ($T=1/\Delta\nu$) & 5000 \\
Number of spectral component ($N+1$) & 46 \\
Number of basis function ($N_{s}$) & 5/10 \\
B-spline order ($k$) & 3 \\
Energy density ($ U_{\mathrm{const}} $) & 150.0 \\ 
\hline \hline
\end{tabular}
\caption{Characterization of the electric field pulse in 1-D.}
\label{tab:pulse1D}
\end{table}

In this section, we consider pair production in 1-D, where we maximize the total number of produced particles along the field direction. The electric field configuration is characterized by the data given in Table~\ref{tab:pulse1D}. We choose a large momentum window $p_3\in[-20,20]$ with a high momentum resolution $\Delta p_3=0.0005$ in order to (i) cover all pairs that are produced along the field direction, and (ii) resolve the fast oscillations in the momentum spectrum. The corresponding number of sampling points is $N_p=80000$.

We individually employ both optimization techniques to identify field configurations that maximize the particle number. For DE, four islands in a ring topology are used with a population of 10 individuals on each island. In the local search algorithm, we selected 10 random initial configurations of which all converged to the same optimal momentum distribution, which indicates that the number of local extrema is possibly small in this case.

\begin{figure}[t]
\centering
\includegraphics[width=0.93\columnwidth]{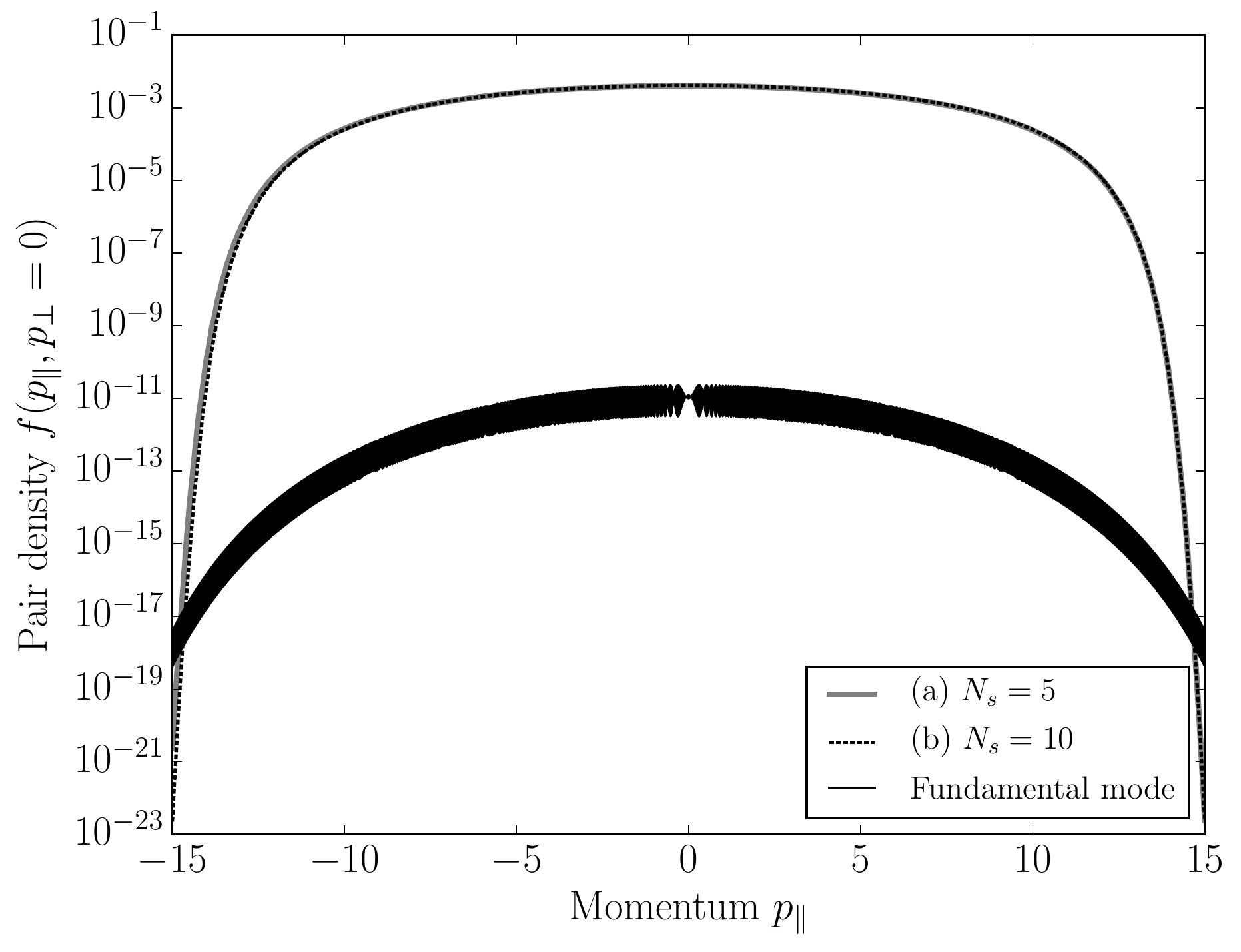}
\caption{1-D momentum spectrum for three different fields: optimal configuration with 5 amplitudes and phases [case (a)], 10 amplitudes with vanishing phases  [case (b)], and a fundamental mode field ($\nu=0.001$). The curve of the fundamental mode spectrum appears thick because of very fast oscillations owing to quantum interferences.}
\label{fig:champion_pair_rate}
\end{figure}

In Fig.~\ref{fig:champion_pair_rate} we display the optimal momentum distribution that has been found individually by both optimization procedures for a different number $N_s$ of basis functions:
\begin{enumerate}
 \item[(a)]{$\vec{a}^{(E)}\in\mathbbm{R}^5$, $\vec{a}^{(\phi)}\in[0,2\pi]^5$ (10--dimensional para\-meter space)}
 \item[(b)]{$\vec{a}^{(E)}\in\mathbbm{R}^{10}$ with fixed $\vec{a}^{(\phi)}\equiv 0$ (10--dimensional parameter space)}
\end{enumerate}
Both choices yield basically indistinguishable momentum spectra that exhibit a smooth Gaussian-like behavior. For comparison we also display the pair density for a single mode (fundamental mode with $\nu = 0.001$) with the same energy density $U_{\mathrm{const}}$ and envelope function. Most notably, the spectrum for optimal momentum distributions are many orders of magnitude above the one for the fundamental mode. Moreover, the fundamental mode spectrum
exhibits fast oscillations, which can be traced back to quantum interferences and are clearly seen as the large line width. This is in stark contrast to the optimal distributions that do not display any fast oscillations. This indicates that optimal field configurations are such that quantum interferences are minimized and suppressed.

\begin{figure}[t]
\centering
\includegraphics[width=\columnwidth]{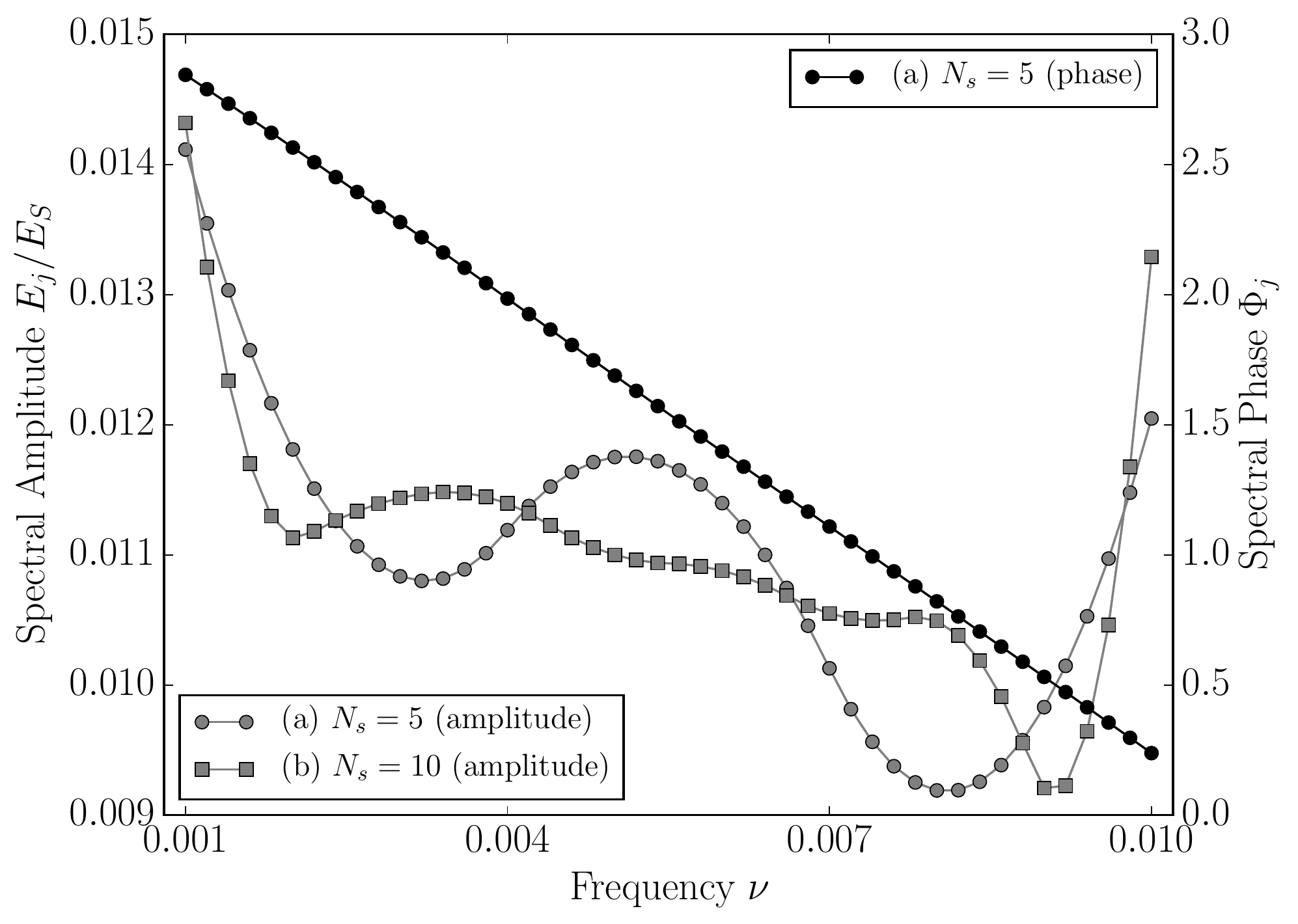}
\caption{Spectral density $\tilde{E}(\omega)$ and spectral phase $\phi(\omega)\sim\omega$ for the optimal field configuration, as function of the frequency $\nu$ in QED units. Despite the fact that the increase of the number of basis functions $B_i(\omega)$ changes the details of $E(\omega)$, we find that the momentum distribution Fig.~\ref{fig:champion_pair_rate} and the electric field Fig.~\ref{fig:champion_efield} remain basically unchanged.}
\label{fig:champion_spectrum}
\end{figure}

\begin{figure}[b!]
\centering
\includegraphics[width=0.95\columnwidth]{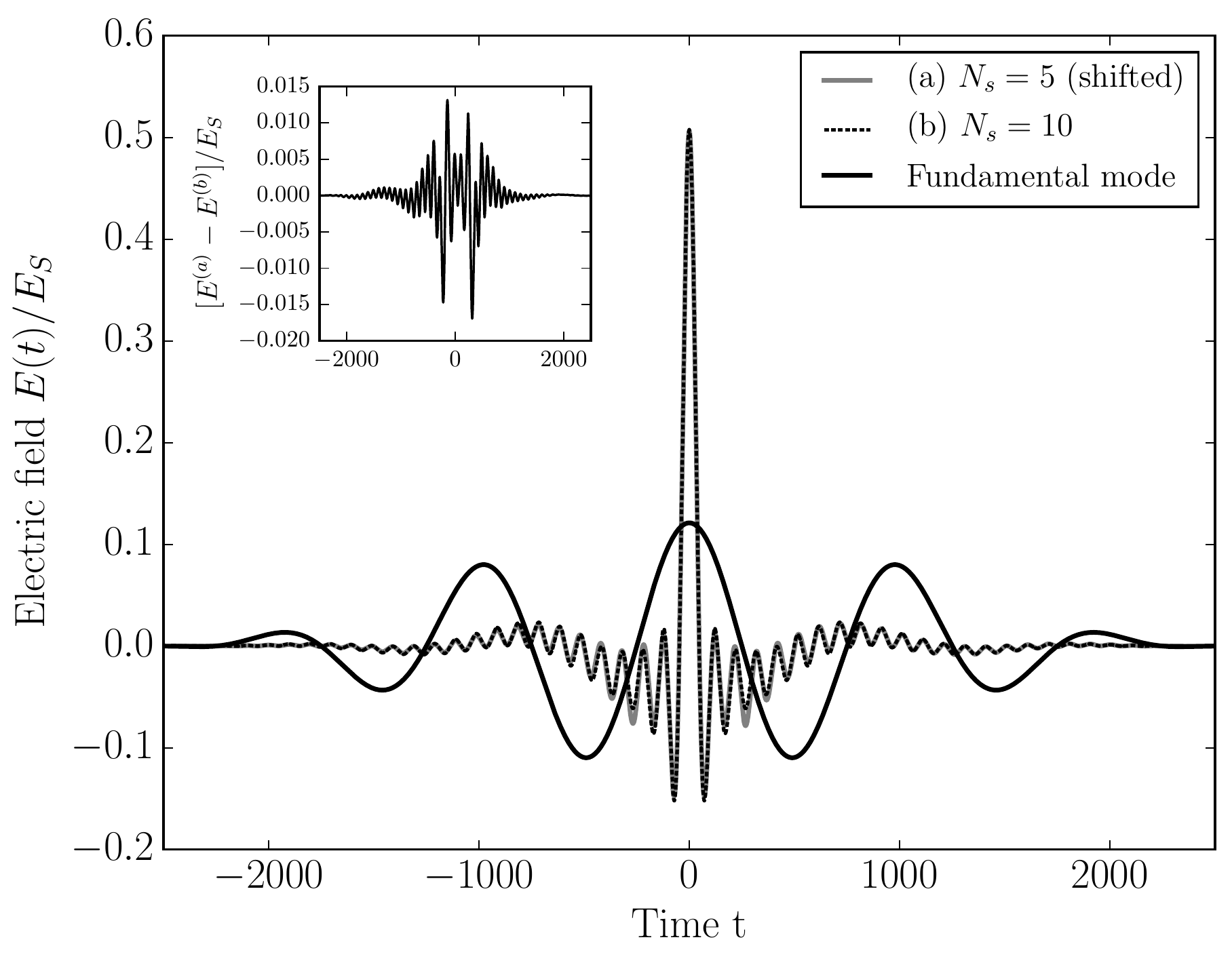}
\caption{Electric field $E(t)$ for the optimal field configuration with $N_s=5$ [case (a)] and $N_s=10$ [case (b)], as function of time $t$ in QED units. For case (a), a time shift of $\Delta t \approx 46.2$ is applied to cancel the phase and to facilitate the comparison between the two cases. The electric field changes only marginally while the spectral density Fig.~\ref{fig:champion_spectrum} shows sizable deviations. We also display the electric field of the fundamental mode in juxtaposition. Finally, the inset contains the difference in the electric field of cases (a) and (b). The maximum difference is approximately 3\% of the maximum field value. Therefore, both electric fields have very similar time dependence. }
\label{fig:champion_efield}
\end{figure}

For case (a), optimizing both amplitudes and phases 
reveals that the optimal field configuration exhibits a linear phase dependence $\phi(\omega)\sim\omega$, corresponding to a time translation of the signal under the envelope, similar to a carrier envelope phase. A typical example of this linear dependence is displayed in Fig.~\ref{fig:champion_spectrum}. On the other hand, the spectral density $\tilde{E}(\omega)$ yields a unique form irrespective of the slope of the spectral phase. In fact, the linear phase dependence is not surprising: as pair production depends exponentially on $|\mathbf{E}|$, the algorithm tends to maximize the field strength. The maximal field strength, however, occurs if all phases are zero since the envelope is centered at $t=0$. In Fig.~\ref{fig:champion_efield} we display the optimal field configuration along with the field of the fundamental mode.

 
Consequently, the solution $\tilde{J}$ with a linear phase dependence yields basically the same momentum spectrum as if all phases are set to zero. In  case (b), 
we hence neglected all phases by setting $\vec{\mathbf{a}}^{(\phi)}=0$ in order to investigate the effect of increasing the number of basis function $B_i(\omega)$. 
The larger number of $N_s$ gives a higher resolution of the spectral density $\tilde{E}(\omega)$, as shown in Fig.~\ref{fig:champion_spectrum}. Accordingly, there are sizable changes in the spectral density. At the same time, however, the optimal momentum distribution Fig.~\ref{fig:champion_pair_rate} and the electric field Fig.~\ref{fig:champion_efield} remain basically unchanged. This indicates that essential features of the optimization problem with $N+1=46$ spectral components are already captured by a B-spline expansion with only $N_s=5$ adjustable parameters. 

\begin{figure}[t]
\centering
\includegraphics[width=\columnwidth]{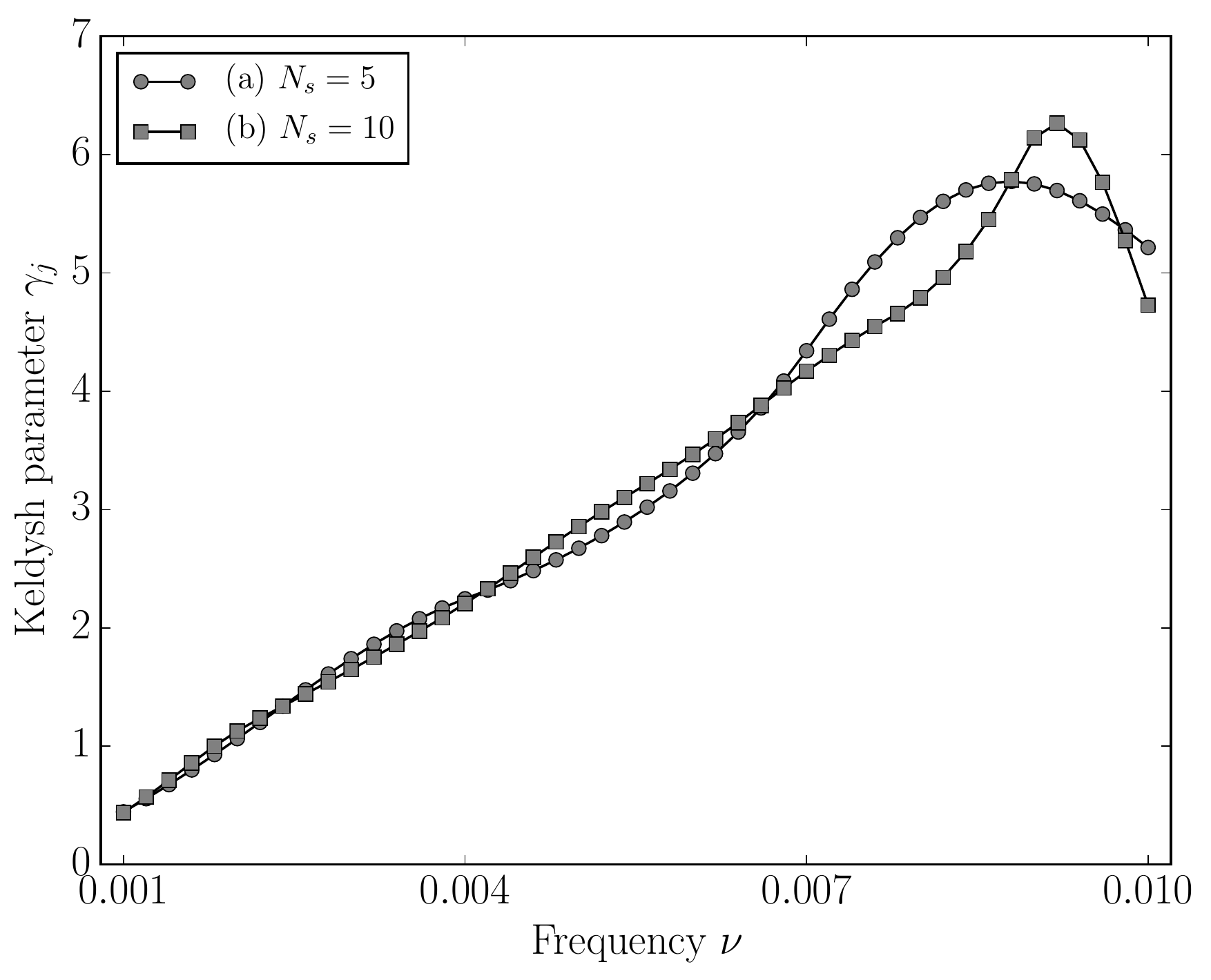}
\caption{Spectral Keldysh parameters $\gamma_{i}$ for the optimal field configuration with $N_s=5$ (case a) and $N_s=10$ (case b), as function of the frequency $\nu$ in QED units. The low-frequency modes are still in the nonperturbative regime whereas the high-frequency modes belong to the multiphoton realm.}
\label{fig:champion_keldysh}
\end{figure}

In Fig.~\ref{fig:champion_keldysh} we display the spectral Keldysh parameter as defined in Eq.~\eqref{eq:keldysh}. We clearly find that some modes are in the nonperturbative regime ($\gamma_{i}<1$)  while others are in the multiphoton regime ($\gamma_{i}>1)$. In this sense, the optimal field configuration exhibits the features of the multimodal dynamically assisted Schwinger mechanism, where the combination of nonperturbative and multiphoton modes enhance pair production. We note again, however, that the meaning of the Keldysh parameter in multiscale problems is not totally clear \cite{PhysRevLett.101.130404,Orthaber:2011cm}.

Our results indicate that there are three essential mechanisms contributing to the maximization of the particle number: (i) Optimal configurations tend to reduce quantum interferences in order to suppress oscillations in the momentum spectrum (see Fig.~\ref{fig:champion_pair_rate}); (ii) Optimal configurations realize the multimodal dynamically assisted Schwinger mechanism where the pair production rate is enhanced by multiphoton modes; (iii) Optimal configurations show a large bandwidth in order to reduce the pulse duration (see Fig.~\ref{fig:champion_spectrum} and Fig.~\ref{fig:champion_efield}). As the integrated energy density $U_{\mathrm{const}}$ is fixed, we obtain higher field strengths and hence a larger number of emitted pairs.

\subsection{Optimization for arbitrary polarization in 2-D}
\label{sec:opt_tot_rate_2D}

This section is devoted to the optimization of pair production for 2-D field configurations. We consider fields of arbitrary polarization (linear, circular, elliptic) and maximize the total number of produced particles in the plane in which the field acts. Table~\ref{tab:pulse2D} summarizes the data that characterizes the laser pulse. 

\begin{table}[t!]
\begin{tabular}{lc}
\hline \hline
Pulse characteristics & Value (QED units) \\
\hline
Minimum frequency ($\nu_{\mathrm{min}} = \omega_{\mathrm{min}}/2\pi$)& 0.02\\
Maximum frequency ($\nu_{\mathrm{max}} = \omega_{\mathrm{max}}/2\pi$)& 0.03\\
$\Delta \nu$ & 0.0002 \\
Pulse length ($T=1/\Delta\nu$) & 5000 \\
Number of spectral component ($N+1$) & 51 \\
Number of basis function ($N_{s}$) & 5/10 \\
B-spline order ($k$) & 3 \\
Energy density ($ U_{\mathrm{const}} $) & 50.0 \\ 
\hline \hline
\end{tabular}
\caption{Characterization of the electric field pulse in 2-D.}
\label{tab:pulse2D}
\end{table}

\begin{figure}[b]
\raggedleft
\includegraphics[width=\columnwidth]{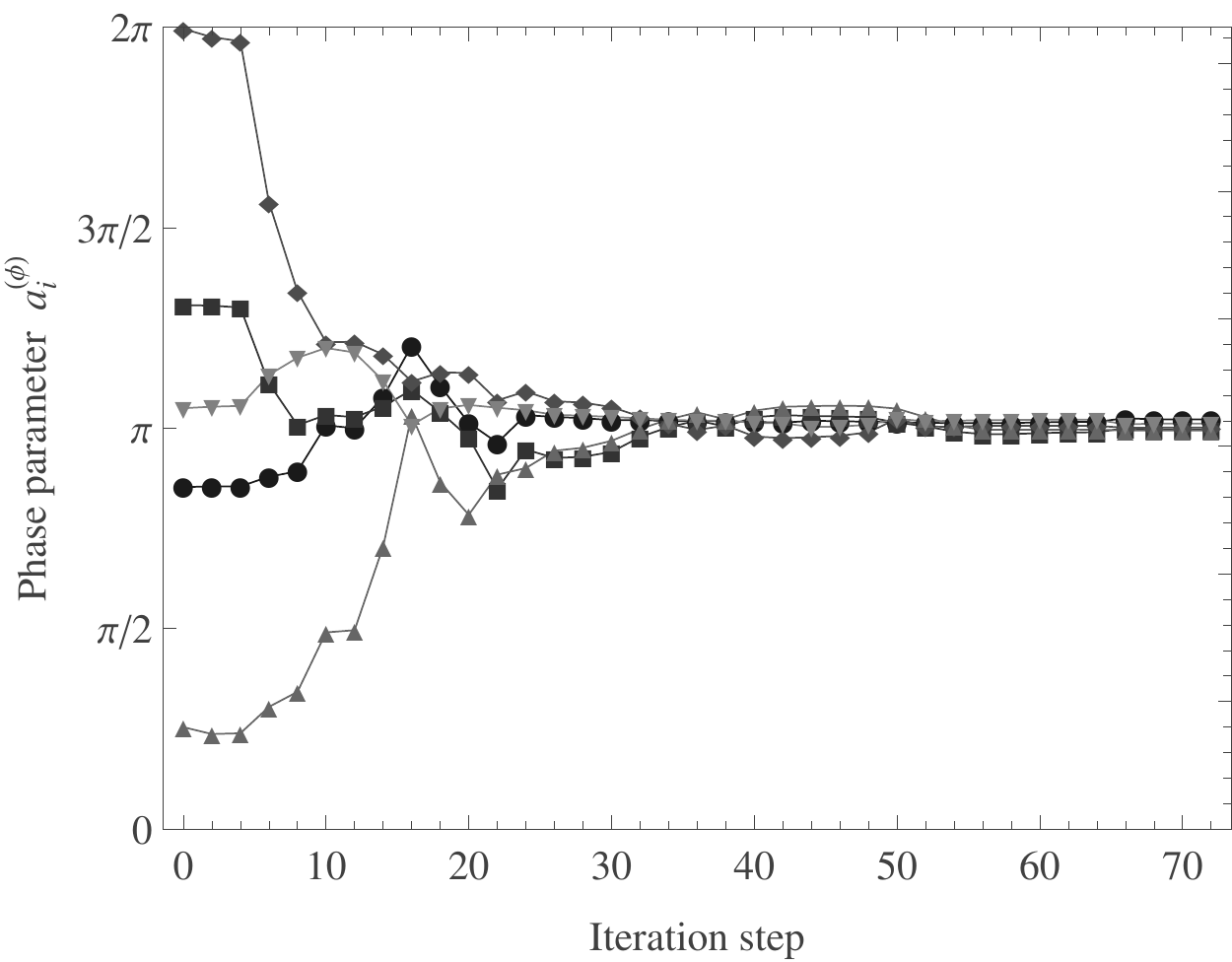}
\caption{Showcase iterative evolution of the phase parameters $\vec{a}_2^{(\phi)}$ using the local search algorithm. Starting from random initial values that correspond to arbitrary ellipticities, the optimal configurations exhibits a linear polarization (relative phase $0$ or $\pi$).}
\label{fig:iterative_evolution}
\end{figure}

In comparison to the 1-D case, we now use higher frequencies $\omega\in[\omega_{\mathrm{min}},\omega_{\mathrm{max}}]$ for computational reasons:  (i) the maximum momenta up to which particles are produced is reduced so that it suffices to choose a smaller momentum window $p_{2,3}\in[-3,3]$ to encompass all produced pairs; (ii) it suffices to choose a lower momentum resolution $\Delta p_{2,3}=0.006$ to resolve the oscillations in the particle spectrum due to quantum interferences. These parameters, however, still account for a much larger number of sampling points $N_p=10^{6}$ which makes computations in 2-D much more expensive than in 1-D. We emphasize, however, that there are no fundamental restrictions for further decreasing the frequency and momentum range given sufficient computational resources. On the other hand, we choose the same optimization parameters for DE (four islands with a population of 10 individuals on each island) and the local search algorithm (10 random initial configurations) as in 1-D.

\textbf{\begin{figure}[b!]
\centering
\includegraphics[width=\columnwidth]{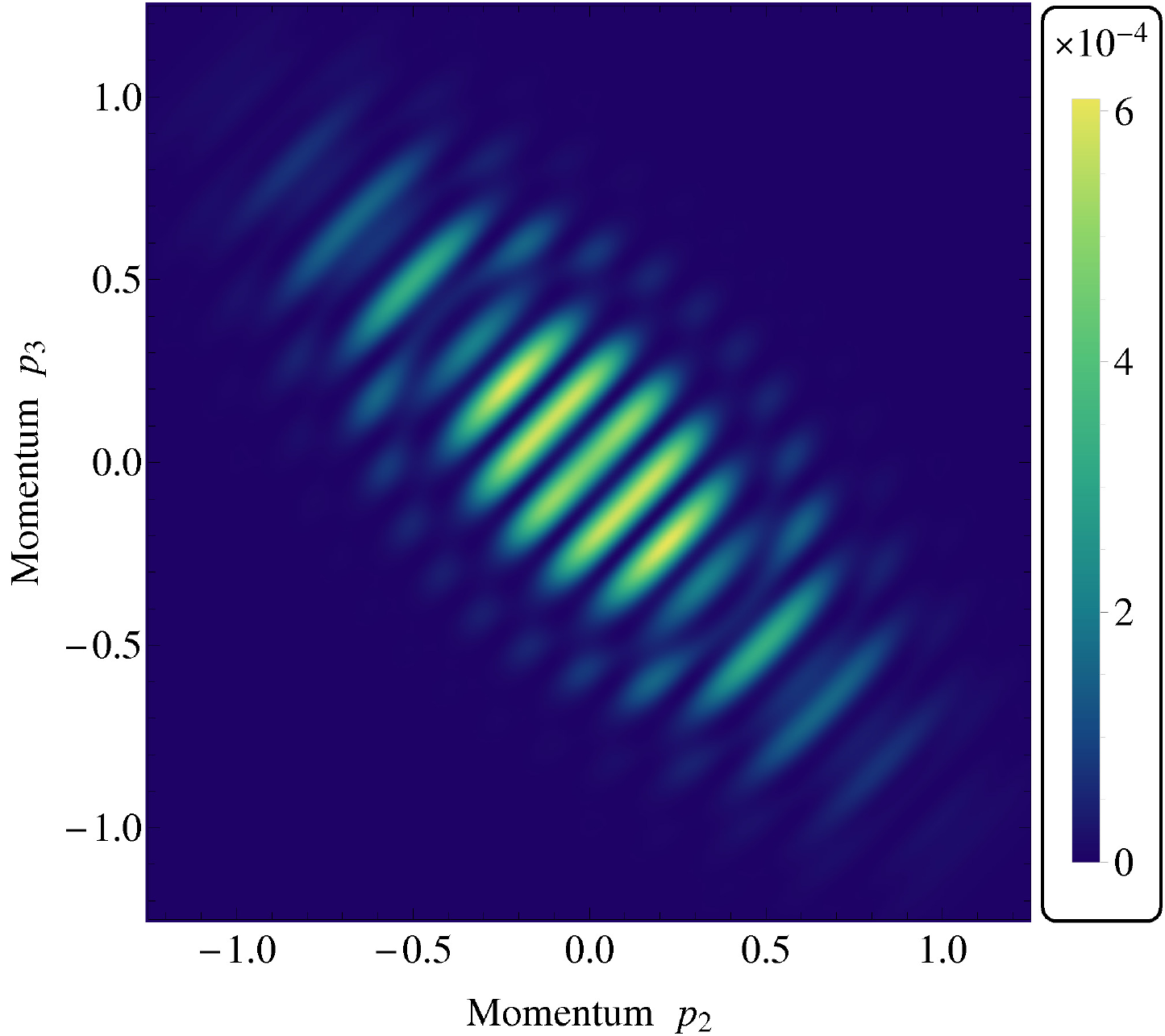}
\caption{[Color online] The 2-D momentum spectrum for the optimal configuration [case (a)] is a linearly polarized electric field. We observe characteristic oscillations along the field direction and a fast decay perpendicular to it.}
\label{fig:2d_champion_linear}
\end{figure}}

We now choose three different instances of optimization parameters to capture all possible polarizations:
\begin{enumerate}
 \item[(a)]{$\vec{a}_{2}^{(E)},\vec{a}_3^{(E)}\in\mathbbm{R}^5$, $\vec{a}_{2}^{(\phi)}\in[0,2\pi]^{5}$, with fixed \mbox{$\vec{a}_{3}^{(\phi)}\equiv 0$} (elliptic polarization, 15--dimensional parameter space)}
 \item[(b)]{$\vec{a}_{2}^{(E)}=\vec{a}_{3}^{(E)}\in\mathbbm{R}^{10}$ with fixed $\vec{\mathbf{a}}^{(\phi)}\equiv 0$ (linear polarization, 10--dimensional parameter space)
 \item[(c)]{$\vec{a}_{2}^{(E)}=\vec{a}_{3}^{(E)}\in\mathbbm{R}^{10}$ with fixed $\vec{a}_{2}^{(\phi)}\equiv 0$, $\vec{a}_3^{(\phi)}\equiv\pi/2$ (circular polarization, 10--dimensional parameter space)}
}
\end{enumerate}
Case (a) corresponds to an optimization problem with an elliptic polarization. Based on the experience from the previous section, we only account for the variation of 5 relative phases which should capture the main features of pair production. We also perform the optimization in the limiting instances of linear polarization [case (b), fixed relative phases $0$] as well as circular polarization [case (c), fixed relative phases $\pi/2$] for a higher number $N_s$ of basis functions.

We first analyze the results of the elliptically polarized field. In Fig.~\ref{fig:iterative_evolution}, we display the iterative evolution of the phase parameters $\vec{a}_2^{(\phi)}$ using the local search algorithm, starting from a random initial configuration. Interestingly, we find that all phases evolve towards the value $\vec{a}_2^{(\phi)}\rightarrow\pi$, corresponding to an electric field that exhibits a linear polarization in each mode. While previous work on short Gaussian pulses with subcycle structure indicated that elliptic or circular polarization might be advantageous at certain high frequencies \cite{PhysRevD.92.085001}, we do not observe this behavior for the multimodal electric field with an energy constraint, as explained below.  

\begin{figure}[b!]
\raggedright
\includegraphics[width=\columnwidth]{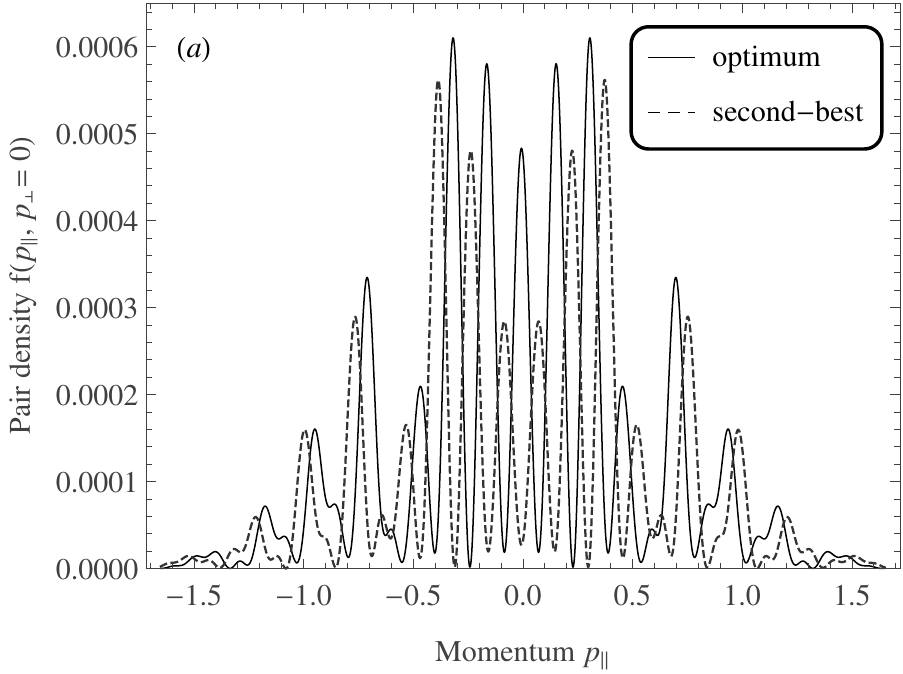}
\\
\vspace{0.35cm}
\centering
\includegraphics[width=\columnwidth]{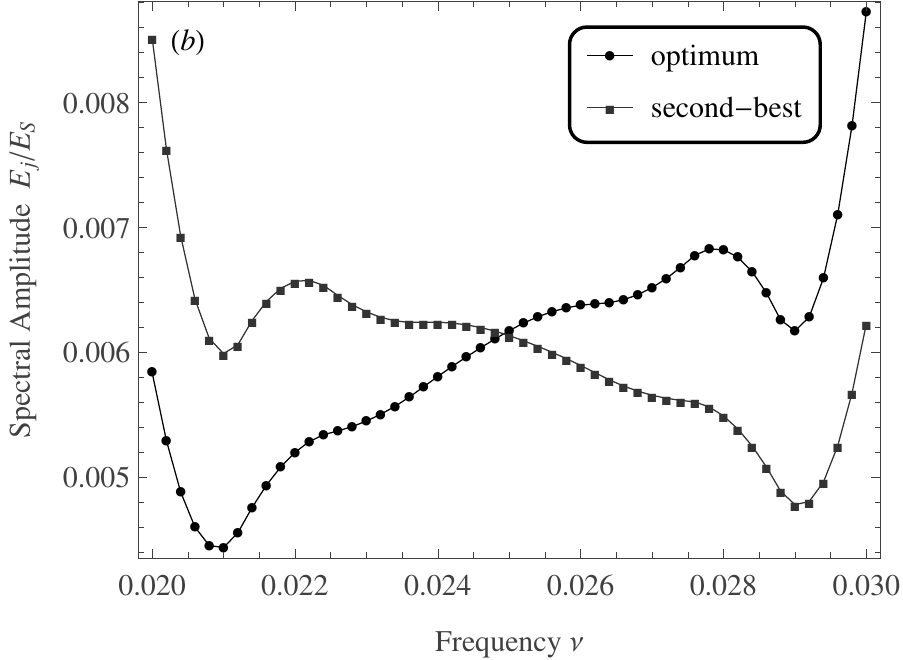}
\caption{(a) Momentum spectra along the linearly polarized electric field for the optimal and the second-best configuration. (b) Spectral amplitudes for the optimal and second-best configuration as function of the frequency $\nu$ in QED units.}
\label{fig:2d_linear_maxima}
\end{figure}

The corresponding optimal momentum distribution, which is found individually by both optimization methods, is displayed in Fig.~\ref{fig:2d_champion_linear}. Unlike the 1-D case (see Fig.~\ref{fig:champion_spectrum}), we now observe oscillations in the spectrum that are typical for time-domain quantum interferences. This is possibly due to the different spectral bandwidth in comparison to the 1-D case, and due to the fact that all spectral components converge to the multiphoton regime with the chosen parameters. Moreover, owing to the alignment of relative phases, the electric field oscillates along the diagonal axis while we still find the typical fast decay of the spectrum along its orthogonal direction, due to the fact that the transverse momentum acts like an effective mass that increases
the gap \cite{PhysRevB.94.125423}. Finally, we note that other equivalent optimal field configurations exist that are connected to the distribution in Fig.~\ref{fig:2d_champion_linear} by a simple rotation in the 2-D plane of the field and pair density. 

To deepen the understanding of the linearly polarized case, we fix the relative phases from the very beginning and increase the number $N_s$ of basis functions [case (b)]. We then find that there are two nearly degenerate local maxima in the total particle number. A cut in their respective momentum spectra along the field direction is displayed in Fig.~\ref{fig:2d_linear_maxima}a. These two configurations mainly differ by their oscillatory behavior: the optimum distribution shows a maximum around zero momentum while the second-best distribution is out of phase there. Although the overall shape of these distributions shows similar features, their spectral amplitudes differ sizeably, as shown in Fig.~\ref{fig:2d_linear_maxima}b. In particular, high-frequency modes dominate for the optimal configuration while the low-frequency modes are enhanced for the second-best configuration. Accordingly, these two configurations are well separated in parameter space.

\begin{figure}[b!]
\centering
\includegraphics[width=\columnwidth]{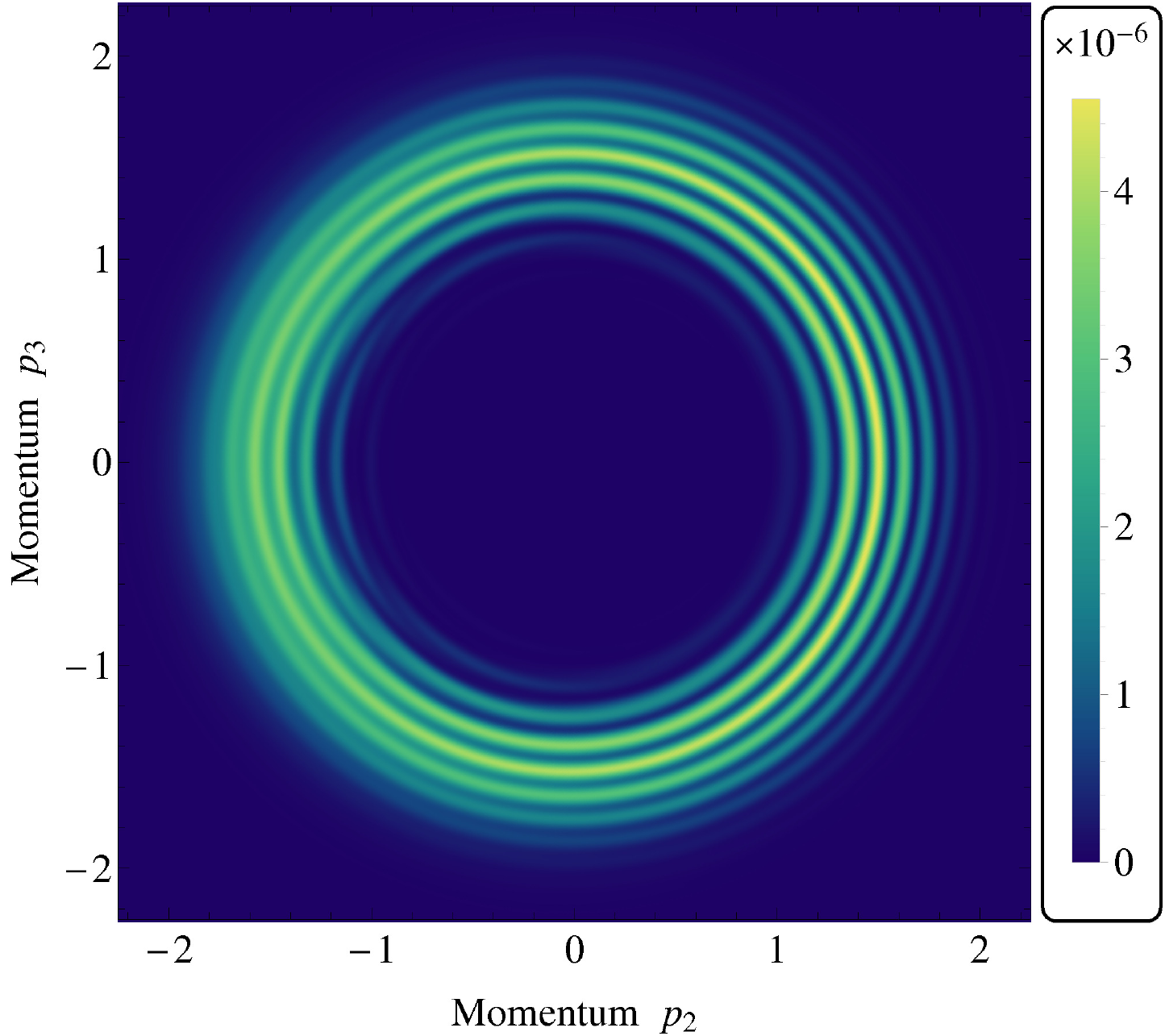}
\caption{[Color online] The 2-D momentum spectrum for the optimal configuration for circularly polarized fields [case (c)]. We observe the characteristic ring structure while no particles are present around zero momentum.}
\label{fig:2d_champion_circular}
\end{figure}

Finally, we also want to understand the absence of circularly polarized modes in the general optimization problem for elliptic polarization. Therefore, we fix again the relative phases from the very beginning and increase the number $N_s$ of basis functions [case (c)]. The corresponding optimum distribution is shown in Fig.~\ref{fig:2d_champion_circular}, where we observe the ring structure that is generic for circularly polarized electric fields \cite{PhysRevD.91.125026}. There are two important characteristics that differ from the linearly polarized case in Fig.~\ref{fig:2d_champion_linear}: First, the maximum value of the distribution function is orders of magnitude smaller in the circularly polarized case. Second, the maximum value is on a ring structure with momentum $|\mathbf{p}|\sim 1.5 $ while it is peaked around zero momentum in the linearly polarized case.

In our understanding, there are two reasons for this behavior: (i) While the produced particles predominantly reside in the vicinity of the origin for the linearly polarized field in case (b), they are expelled to much higher momentum for circular polarization in case (c). Accordingly, it seems that a large part of the energy is used for this acceleration rather than for particle production; (ii) the peak electric field strength ratio between the linear and circular case is approximately $E_{\mathrm{lin}}^{\mathrm{peak}}/E_{\mathrm{circ}}^{\mathrm{peak}}\sim\sqrt{2}$. This is attributed to the fact that we are fixing the integrated energy density to a given value $U_{\mathrm{const}}$, which can easily be checked for a single mode. Since the particle production depends exponentially on the field strength, the linear configuration is superior in the number of generated electron-positron pairs. 

\subsection{Performance of the numerical methods}
\label{sec:perf_num}

We now analyze the performance of the numerical methods that were used to obtain the results given in previous sections. The optimization techniques are compared first, followed by an analysis of the pair production techniques.

\subsubsection{Optimization techniques}

The comparison of performance between local search (see Sec.~\ref{sec:opt_local}) and DE (see Sec.~\ref{sec:opt_meta}) can be carried out by counting the number of objective function evaluations that are required to reach a converged solution. This number can fluctuate substantially from one calculation to the other as some steps in both algorithms are random. In particular, initial conditions are chosen randomly so  that they can be far from a maximum in parameter space or in a region where the gradient is very small. Both are detrimental to the performance of optimization algorithms. Moreover, the crossover and mutation in DE involve random operations which may select a slower evolution path. Finally, the parameter space dimension also has a direct influence on the convergence of optimization methods: larger parameter spaces obviously demand more objective function evaluations.

As a consequence, a rigorous comparison of the techniques used in this article goes beyond the scope of this work. Nevertheless, our calculation sample allows us to infer some rough tendencies. In the current example with 10 random initial conditions, the local search method required on average $\approx 80-100$ objective function evaluations per initial condition to converge to a local maximum, i.e., $\approx 800-1000$ to identify the global maximum. In contrast, the DE needed on average $\approx 900-1300$ objective function evaluations to converge to the global maximum in parameter space. 

We hence conclude that the two methods perform similarly on the current problem (note, however, that an update step in the local search method is twice as expensive as in the DE since the equations of motion need to be evolved forward and backward in time to compute the local gradient). The fact that a comparatively small number of random initial condition suffices in the local search algorithm in order to identify a single maximum hints at a small number of local extrema in the current problem. We expect, however, that the DE outperforms the local search algorithm for cost functions that exhibit a large number of local extrema since a higher number of random initial conditions would be required to safely identify the global maximum.

\subsubsection{Pair production calculation techniques}

We also compare the computational performance of the Dirac equation approach (see Sec.~\ref{sec:DE}) to the kinetic formulation (see Sec.~\ref{sec:QKE}). This test is performed by looking at the numerical error in the pair density
\begin{align}
\epsilon := \frac{\int_{\mathcal{D}_{\mathbf{p}}} d^{3}p |f(t_{f},\mathbf{p})-f_{\mathrm{exact}}(t_{f},\mathbf{p})|}{\int_{\mathcal{D}_{\mathbf{p}}} d^{3}p\,f_{\mathrm{exact}}(t_{f},\mathbf{p})} \ ,
\end{align} 
as the number of time steps $N_{t}$ is increased.
Here, $f_{\mathrm{exact}}(t_f,\mathbf{p})$ is the exact distribution function whereas $f(t_f,\mathbf{p})$ denotes the numerically determined approximate solution. Obviously, the computation time grows as $N_t$ is increased for a fixed evolution time $T:=t_f-t_i$. At the same time, however, the numerical error in the ODE solvers decreases polynomially like $\mathcal{O}(N_{t}^{-q})$, where $q$ is the order of convergence. In the current study, we used second-order split-operator method in the Dirac equation approach ($q=2$) and a fourth-order Runge-Kutta method for the quantum kinetic equation ($q=4$). In order to compare the two methods, we compute the numerical error $\epsilon$ for the optimal circularly polarized field configuration that was found in Sec.~\ref{sec:opt_tot_rate_2D} upon variation of $N_{t} \in [2\cdot 10^{4},10^{6}]$. Following a standard procedure in numerical analysis, $f_{\mathrm{exact}}$ is obtained by a solution with a large number of time steps (here, we choose $N_{t}=2\cdot 10^{6}$ which yields a solution on the level of machine precision). 

\begin{figure}[t]
\centering
\includegraphics[width=1.09\columnwidth]{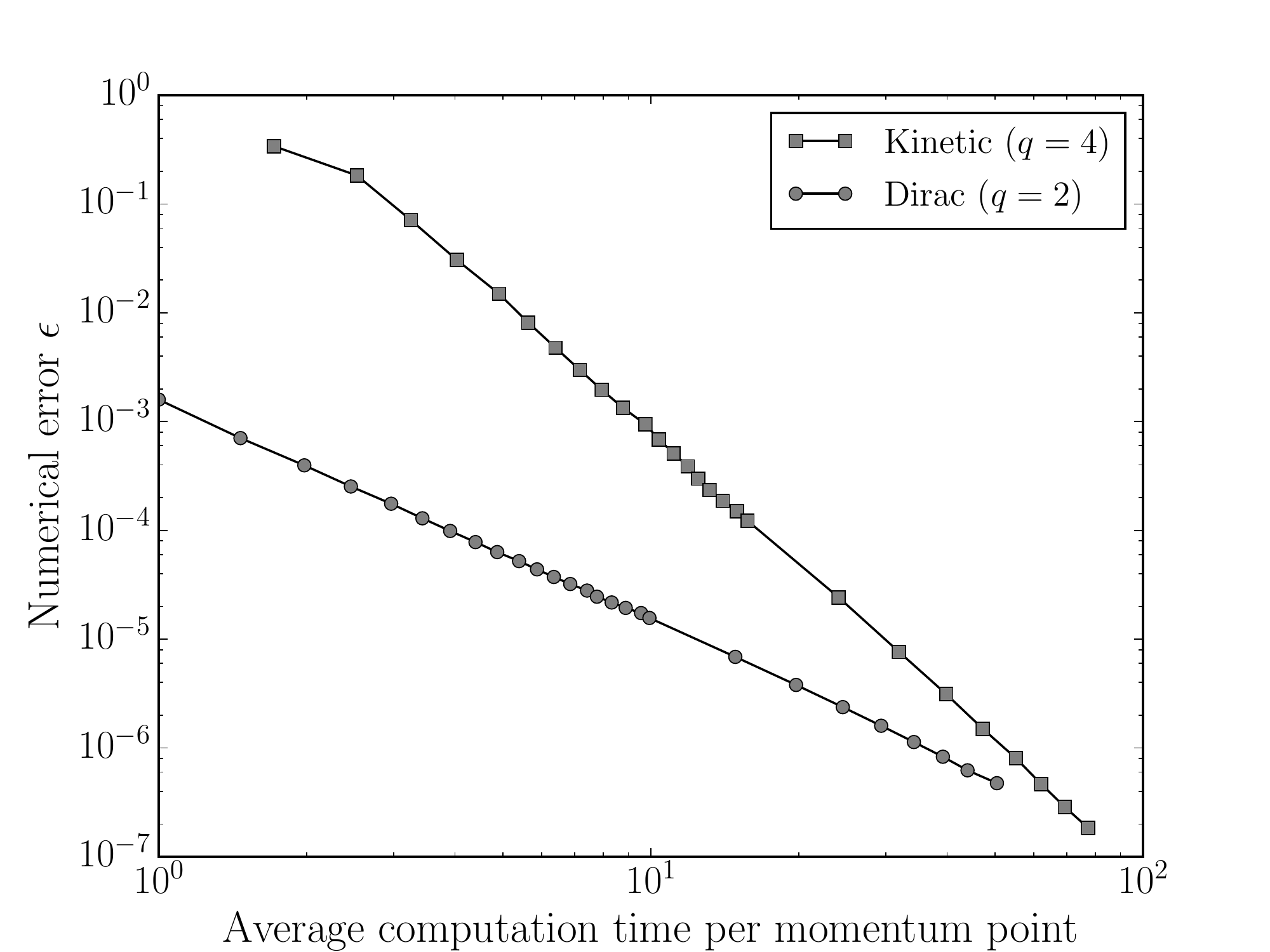}
\caption{Numerical error of both numerical methods as function of the average computation time per momentum point. The computation time is normalized by the smallest measured time, which corresponds to the Dirac method with $N_t=2\cdot 10^4$.}
\label{fig:pair_rate_accuracy}
\end{figure}

In Fig.~\ref{fig:pair_rate_accuracy} we display the numerical error as a function of the average computation time per momentum point for both computation methods. These results indicate a clear advantage of the Dirac method in comparison to the kinetic formulation in the regime of small values of $N_{t}$. For instance, it suffices to take $N_t\approx 2\cdot10^{4}$ in the split-operator method while it requires a much larger value $N_t\approx10^{5}$ in the kinetic formulation in order to obtain a numerical error at the level of $\epsilon=10^{-3}$. Consequently, we observe a gain of a factor of $\approx 5$ in computational performance at this error level. We surmise that this remarkable success of the Dirac method is due to the inherent unitarity conservation of the split-operator method, which allows for fine cancellations when the wave function is projected over the positive energy state. One clearly sees, however, that the kinetic method catches up with the Dirac method for large values of $N_{t}$ owing to the higher convergence order.

We note that the performance of both methods can be further improved in principle by resorting to more accurate numerical schemes. For instance, a splitting scheme with fourth order accuracy could be used to solve the Dirac equation or higher order Runge-Kutta techniques could be applied to solve the ODE system in the kinetic formulation. These matters are left for future investigations.  

\section{Conclusion}
\label{sec:conclu}

In this work, we performed a pulse shape optimization procedure on multimodal homogeneous electric fields with arbitrary polarization to maximize electron-positron pair production. Our results demonstrate that the multimodal character of the field is  essential for
the maximization of the pair density. As a matter of fact, all the optimized pulses  that were found with our techniques have non-vanishing spectral amplitudes over the whole bandwidth. 
As a consequence, multimodal fields offer enough control for canceling and controlling the emergence of interference effects so that the resulting pair densities exhibit less oscillations.
Moreover, having a large bandwidth reduces the pulse duration and compresses the available energy into a short time interval, making for pulses with higher field strengths. Finally, it allows for the possibility of having certain spectral components in the Schwinger and multiphoton regimes, realizing the multimodal dynamically assisted Schwinger mechanism. All of these mechanisms conspire to increase the pair production rate.   

We also studied the effect of the field polarization on electron-positron pair production. For a given integrated energy density of the laser pulse, our results indicate that linearly polarized electric field are superior over circular or elliptically polarized ones. These findings could be useful for experimental attempts to detect the Schwinger mechanism at future high-intensity laser facilities such as the Extreme Light Infrastructure (ELI) or the Exawatt Center for Extreme Light Studies (XCELS). 

In this work, we substantially extended previous investigations on the optimization of linearly polarized electric fields \cite{PhysRevD.88.045028, Hebenstreit2014189,Hebenstreit2016336} in several ways: 

(i) We parametrized the field configurations in Fourier space by using a polynomial basis expansion. Using this approach, we showed that the number of parameters that is required to characterize optimal field configurations is much smaller than the number of field modes itself. In fact, we found that corrections due to the increase of parameter space are only of the order of a few percent. Accordingly, we were able to efficiently optimize multimodal electric field with up to $50$ Fourier modes on much smaller parameter spaces. 

(ii) To this end, we employed both a local search algorithm as well as population based metaheuristics. We found that both methods have similar computational performance in the current problem, measured by the number of objective function evaluations. This is largely due to the fact that the objective function seems to have only a small number of local extrema. However, we expect that metaheuristic algorithms outperform the local search method once objective functions exhibiting a larger number of local extrema are considered, such as e.g., in the inverse problem of Schwinger pair production \cite{Hebenstreit2016336}. For electron-positron pair production, we were able to safely identify a single maximum with the metaheuristic algorithm and with the local search algorithm by taking only 10 random initial conditions. Although there is no rigorous proof that this is the global maximum, the fact that both methods converge towards the same individual indicates that there is a significant probability that we found the optimal solution in the whole parameter space.  

(iii) In order to optimize particle production in two-dimensional electric fields, it was necessary to use efficient numerical schemes to compute the momentum spectrum: one needs to solve repeatedly the dynamic equations on $\mathcal{O}(10^6)$ sampling points in order to resolve the two-dimensional momentum space. On the one hand, we derived the quantum kinetic equations based on the DHW phase space approach and solved them using a fourth-order Runge-Kutta method. To this end, we explicitly showed how to lift the redundancy of the previously employed formalism (10 coupled ODEs) \cite{PhysRevD.89.085001} in order to obtain a system of only 6 coupled ODEs. This formalism can be considered as the generalization of the well-known quantum kinetic equation for linearly polarized fields \cite{PhysRevD.60.116011}. On the other hand, we directly solved the Dirac equation by using a split-operator method with second order convergence. This method turned out to be the more efficient approach, which showed a significant gain in computational performance in comparison to the quantum kinetic formulation.

The techniques that were developed and used in this work may also be beneficial for different problems, such as the reduction of the number of pairs for the shortcut to adiabaticity \cite{1367-2630-18-1-012001}, which may be important for particle-hole generation in condensed matter systems, or the inverse problem of Schwinger pair production \cite{Hebenstreit2016336}. Actually, the employed methods are expected to be applicable for any system that is well-described by the Dirac equation. On the other hand, the field parametrization is not restricted to relativistic quantum mechanics. In fact, it could be useful for the control of other physical systems, such as the generation of harmonics in atomic and molecular physics, as long as the system under consideration is coupled to a spatially homogeneous but time-varying electric field. 

To conclude, we made an attempt to (i) design field configurations that approach realistic experimental setups as much as possible, and (ii) push the computational boundaries to investigate the pair production process. To this end, we introduced electric field configurations with a large bandwidth that were parameterized in Fourier space. In fact, the large bandwidth allows for short pulses with high field strength, which are typical in experimental proposals for the detection of the Schwinger mechanism, while the numerical methods allowed us to consider comparatively long (few-attosecond) pulses and push the computational boundary. Still, our investigation fell short on some aspects: we considered laser pulses that have much higher field strength and much shorter pulse duration than those which can be achieved experimentally with current technology. Moreover, unlike realistic tightly focused laser beams, we neglected any spatial dependence and the effect of magnetic fields. Nevertheless, our study still exhibits important trends that are supposed to hold even if these effects are taken into account, which is a crucial issue for future investigations.    

\begin{acknowledgments}
The authors would like to thank P.~Blain for coding a part of the Dirac solver and J.~Dumont, C.~Lefebvre, E.~Lorin and A.~D.~Bandrauk for discussions. Computations were performed on UBELIX, the HPC cluster at the University of Bern, and on the supercomputer MAMMOUTH from Universit\'{e} de Sherbrooke, managed by Calcul Qu\'{e}bec and Compute Canada. The operation of this supercomputer is funded by the Canada Foundation for Innovation (CFI), the minist\`{e}re de l'\'{E}conomie, de la science et de l'innovation du Qu\'{e}bec (MESI) and the Fonds de recherche du Qu\'{e}bec -- Nature et technologies (FRQNT). This research was supported in part by the National Science Foundation under Grant No. NSF PHY11-25915 and by the European Research Council under the European Union's Seventh Framework Programme (FP7/2007-2013)/ERC under grant agreement 339220.
\end{acknowledgments}

\bibliographystyle{apsrev4-1_mod}
\bibliography{bibliography}

\end{document}